\documentclass[twocolumn]{aa}

\usepackage{color}
\usepackage{epsfig}
\usepackage{soul}
\usepackage{float}
\usepackage{enumitem}
\usepackage{appendix}
\usepackage{wrapfig}
\usepackage{graphicx}
\usepackage{multicol}
\usepackage{hyperref}
\usepackage{chemfig}
\usepackage{longtable}
\usepackage{supertabular,booktabs}
\usepackage{pdflscape}
\usepackage{longtable}
\definecolor{NGC1333}{HTML}{006D77}
\definecolor{B1}{HTML}{5C3D73}
\definecolor{IC348}{HTML}{922D50}
\definecolor{B5}{HTML}{DAA520}

\usepackage[version=3]{mhchem}
\hypersetup{
    colorlinks=true,
    linkcolor=blue,
    filecolor=magenta,
    citecolor=blue,
    urlcolor=blue,
    pdftitle={Overleaf Example},
    pdfpagemode=FullScreen,
    }
\raggedcolumns
\usepackage{natbib}

\def\aap{A\&A}
\def\apjs{ApJS}

\usepackage{blindtext}
\usepackage{lipsum}
\usepackage{afterpage}

\newsavebox{\splitcaptionbox}
\newlength{\splitcaptionheight}
\makeatletter
\long\def\@makesplitcaption#1#2{
  \if@stage@final
    \vskip0.7\abovecaptionskip
    \addtolength{\splitcaptionheight}{-0.7\abovecaptionskip}%
  \else
    \vskip\abovecaptionskip\goodbreak
    \addtolength{\splitcaptionheight}{-\abovecaptionskip}%
  \fi
  \setbox\splitcaptionbox=\vbox{\interlinepenalty 0
    \reset@font\small{\bfseries#1.} #2\par}%
  \vsplit\splitcaptionbox to \splitcaptionheight
  \global\setbox\splitcaptionbox=\box\splitcaptionbox
  \if@cop@home\ifonline\ifnum\csname c@\@captype\endcsname=1 
    \immediate\write\@auxout{\string\gdef\string\@num\@captype{}}%
    \hypertarget{\@captype}{}%
  \fi\fi\fi}

\newcommand{\splitcaption}[3][\empty]
 {\setlength{\splitcaptionheight}{#3}%
  \let\cop@makecaption=\@makecaption
  \let\@makecaption=\@makesplitcaption
  \ifx\empty#1\relax
    \caption{#2}%
  \else
    \caption[#1]{#2}%
  \fi
  \let\@makecaption=\cop@makecaption}
\makeatother

\newcommand{\mergecaption}{\ifdim\ht\splitcaptionbox>0pt
  \begin{figure}[tp]
  \box\splitcaptionbox
  \end{figure}
\fi}

\begin{document}

\title{c-C$_{3}$H$_{2}$ deuteration towards prestellar and starless cores in the Perseus Molecular Cloud}

\author{J.~Ferrer Asensio, \inst{1} S.~Scibelli,\inst{2}, \thanks{Jansky Fellow of the National Radio Astronomy Observatory.} L.~Steffes,\inst{5} B.~Kulterer,\inst{3} A.~Pokorny-Yadav,\inst{2,4} Y.~Shirley,\inst{5} A.~Meg{\'i}as,\inst{6} I.~Jim{\'e}nez-Serra,\inst{6} A.~Taillard\inst{6} }
\institute{\tiny{\inst{1} RIKEN Cluster for Pioneering Research, Wako-shi, Saitama, 351-0106, Japan\\} \inst{2} National Radio Astronomy Observatory, 520 Edgemont Road, Charlottesville, VA 22903, USA\\ \inst{3} Department of Astronomy, University of Virginia, 530 McCormick Rd, Charlottesville, Virginia 22904, USA \\ \inst{4} Department of Astronomy, University of California, Berkeley, Berkeley, CA, 94720, USA \\ \inst{5} Steward Observatory, University of Arizona, 933 North Cherry Avenue, Tucson, AZ 85721, USA\\ \inst{6} Centro de Astrobiolog\'ia (CAB), CSIC-INTA, Carretera de Ajalvir, km 4, E-28805 Torrej\'on de Ardoz, Spain \\}

\titlerunning{c-C$_{3}$H$_{2}$ deuteration towards prestellar and starless cores in the Perseus Molecular Cloud}
\authorrunning{J. Ferrer Asensio et al.}

\date{Received  ; accepted }

\abstract
{In cores deuterium fractionation becomes highly efficient due to low temperatures and CO freeze-out. Cyclopropenylidene (c-C$_{3}$H$_{2}$), a small cyclic molecule formed early in chemical evolution, and its deuterated forms serve as valuable tracers of gas-phase deuteration in these environments.}
{In order to statistically explore the c-C$_{3}$H$_{2}$ deuteration ratios towards starless and prestellar cores, we present observations of c-C$_{3}$H$_{2}$ and its deuterated isotopologues on a sample of cores in the Perseus Molecular Cloud. }
{Transitions of c-C$_{3}$H$_{2}$, c-C$_{3}$HD and c-C$_{3}$D$_{2}$ were observed with the Yebes 40m, the Arizona Radio Observatory (ARO) 12m and the Institut de Radioastronomie Millimétrique (IRAM) 30 meter telescopes towards a total of 16 starless and prestellar cores in the Perseus Molecular Cloud. The lines were fitted with Gaussian profiles and their column densities were computed using the non-LTE software RADEX.}
{The main isotopologue c-C$_{3}$H$_{2}$ is detected in 93\% (14/15) of the targeted cores (for one of the 16 cores none of its transitions were covered), its singly-deuterated form c-C$_{3}$HD is detected in 94\% (15/16) of the targeted cores and its doubly-deuterated form c-C$_{3}$D$_{2}$ is detected towards 56\% (9/16) of the cores detected. A range of column densities towards the different cores was derived to be: for c-C$_{3}$H$_{2}$, (0.5 - 8.1)$\times$10$^{13}$ cm$^{-2}$; for c-C$_{3}$HD, (0.2 - 2.1)$\times$10$^{12}$ cm$^{-2}$ and for c-C$_{3}$D$_{2}$, (0.6 - 1.6)$\times$10$^{11}$ cm$^{-2}$. The ortho-to-para ratio of c-C$_{3}$H$_{2}$ was obtained for all except one core with a median value of 3.5 $\pm$ 0.4. The D/H and D$_{2}$/D ratios were obtained for the cores with detections, yielding a statistically corrected D/H range of 0.5 - 9.2\% with a median value of 1.5 $\pm$ 0.2 \% and a statistically corrected D$_{2}$/D range  9.0 - 55.2\% with a median value of 25.9 $\pm$ 4.3 \%.}
{No apparent trend is seen with ortho-to-para ratio of c-C$_{3}$H$_{2}$ and the evolutionary stage of the core, as traced by volume density $n_{\rm H_2}$. The median c-C$_{3}$H$_{2}$ D/H ratio in Perseus' starless cores appears lower than the value for the Taurus Molecular Cloud and Chamaeleon Molecular Cloud. The D$_{2}$/D ratio is equivalent between the Perseus and Taurus Molecular Clouds within uncertainties. There is a correlation with the D/H ratio and the $n_{\rm H_2}$ of the cores in Perseus, strengthening the idea of D/H being a tracer of core evolution. The D$_{2}$/D ratio does not correlate with $n_{\rm H_2}$, but positively correlates with $T_{\rm kin}$, suggesting that its formation is favoured by a slightly endothermic reaction. }


\maketitle

\section{Introduction} \label{introduction3}

The deuterium to hydrogen ratio (D/H) in the local interstellar medium (ISM) is 2.0 $\pm$ 0.1 $\times$ 10$^{-5}$ \citep{linsky:03, prodanovic:10, caselli:12, ceccarelli:14}. Locally, molecules can be enriched with deuterium resulting in higher D/H ratios compared to the value observed in the local ISM. This process, known as deuterium fractionation, happens efficiently in prestellar cores. These cores, that represent the earliest stage of the star formation process, are gravitationally bound objects which show signs of contraction motions and are characterised by high volume densities ($n_{\rm H_2}$ $\sim$ 10$^{6}$ cm$^{-3}$) and low temperatures (\textit{T} $<$ 10 K) at their centres \citep{bergin:07, keto:08}. These are a subgroup of starless cores, which are also defined as gravitationally bound objects, but do not show signs of contraction motions. The low temperatures favour the formation of bonds with deuterium over hydrogen due to the lower zero point energy (ZPE) of deuterium. Moreover, the two-way reaction involving one of the main deuteration agents in the gas phase, H$_{2}$D$^{+}$: \begin{equation}
    {\rm H_3^+ + HD \rightleftharpoons H_2D^+ + p \text{-} H_2 + \Delta E}
\end{equation} is shifted to the right hand side because at low temperatures the backwards reaction is quenched \citep{pagani:92}. This is due to the exothermicity of the reaction and the H$_{2}$ ortho-to-para (otp) ratio being low in prestellar cores ($<$ 0.01; \citealp{pagani:09, dislaire:12}) compared to its statistical value of 3. The higher energy of ortho-H$_{2}$ with respect to para-H$_{2}$ can overcome the reaction barrier destroying H$_{2}$D$^{+}$ in favour of H$_{3}^{+}$. Furthermore, as the temperatures decrease towards the centre of the core and with the contraction of the core, the main destructor of H$_{2}$D$^{+}$, CO, is frozen onto the surface of dust grains. All of these conditions increase the H$_{2}$D$^{+}$/H$_{3}^{+}$ ratio. Then, the H$_{2}$D$^{+}$ can transfer the deuterium atom to other molecules. \

Prestellar cores contract with time, which lowers the temperature at the centre as well as increases the area of CO freeze out. Thus, colder and longer-lived prestellar cores will present higher deuterium fractionation ratios as well as multi-deuterated molecules compared to warmer and shorter-lived ones. This was inferred from modelling methanol (CH$_{3}$OH) deuteration in prestellar cores using observed D/H and D$_{2}$/D ratios towards low and high-mass protostellar sources \citep{vangelder:22}. Thus, the D/H and D$_{2}$/D molecular ratios are a great tool to better understand the characteristics of the prestellar core nature.

Nevertheless, because of the low temperatures of starless and prestellar cores, studying the deuteration of molecules larger than methanol is challenging due to low abundances. Even doubly deuterated methanol has only been reported in the literature for three cores \citep{lin:23, scibelli:25}. This is why, in this work we explore the deuterium fractionation in starless and pre-stellar cores by focusing on the deuteration of a smaller molecule, cyclopropenylidene (c-C$_{3}$H$_{2}$) towards the Perseus Molecular Cloud. \

c-C$_{3}$H$_{2}$ is a small organic molecule detected towards multiple sources at different stages of the star formation process such as diffuse clouds, starless and prestellar cores, circumstellar envelopes, hot corinos, planetary nebulae, and outflow cavities \citep{thaddeus:85, cox:88, madden:89, lucas:00, martinezhenares:25}. Its deuterated forms have also been detected towards the interstellar medium \citep{bell:86, gerin:87, spezzano:13, majumdar:17}. Deuterated cyclopropenylidene has only been detected towards a handful of prestellar cores in Taurus: \cite{gerin:87, spezzano:13, gratier:16, chantzos:18, giers:22} and in Chamaeleon: \cite{lis:25}. c-C$_{3}$H$_{2}$ is mainly formed by the radiative recombination of c-C$_{3}$H$_{3}^{+}$ in the gas phase \citep{loison:17}. c-C$_{3}$HD and c-C$_{3}$D$_{2}$ are formed from the successive deuteration of the main isotopologue, also in the gas phase \citep{spezzano:13}. The singly- and doubly-deuterated c-C$_{3}$H$_{2}$ abundances towards the L1544 prestellar core can be reproduced by chemical models only using gas-phase reactions \citep{spezzano:13}. This makes c-C$_{3}$H$_{2}$ a unique tracer for gas-phase only deuteration in contrast to, for example methanol, which is thought to form uniquely on the surface of grains \citep{watanabe:02, osamura:04, watanabe:05, fuchs:09, hidaka:09}.\

Perseus is one of the nearest ($\sim$294 $\pm$ 17~pc; \citealt{zucker:18}) and most extensively studied molecular clouds in the solar neighbourhood. Covering a total area of about 74~pc$^{2}$ \citep{evans:09}, the cloud extends roughly 10~pc across the sky and exhibits a velocity gradient with local standard of rest (LSR) velocities ($v_{\rm LSR}$) ranging from 4.5 to 10~km~s$^{-1}$, possibly indicating the superposition of multiple cloud components \citep{arce:10}. Perseus is a low-mass star-forming region that hosts a rich population of young stellar objects (YSOs), including about 100 dense cores and over 400 YSOs, of which approximately 50 are classified as Class~0 and Class~I protostars \citep{yang:21}. The cloud contains two main protostellar clusters, NGC~1333 and IC~348, along with other regions of active star formation such as B5, B1, L1448, and L1455. This region harbours a large number of low-mass pre-main sequence stars, embedded protostars, and starless or prestellar cores \citep[e.g.,][]{ladd:93, aspin:94, lada:95, hatchell:05, enoch:06, kirk:06, muench:07, gutermuth:08, evans:09}. In particular, NGC~1333 exhibits numerous active molecular outflows, making it one of the most dynamically young and active subregions within the cloud. Recent chemical surveys of embedded protostars in Perseus, conducted with the Nobeyama telescope, have detected species such as C$_2$H, c-C$_3$H$_2$, and CH$_3$OH, revealing possible correlations between source location and the CH$_3$OH/C$_2$H ratio, which may reflect environmental influences on the chemical composition of these protostellar systems \citep{higuchi:18}. \citet{scibelli:24} surveyed 35 starless and prestellar cores in the Perseus Molecular Cloud with the ARO~12~m telescope, detecting methanol (CH$_3$OH) and acetaldehyde (CH$_3$CHO) in 100\% and 49\% of the sample, respectively. Follow-up Yebes~40~m observations revealed several additional complex organic molecules (COMs), showing that such species are already widespread in the cold gas preceding star and planet formation in Perseus. In this work we continue the chemical survey of the starless and prestellar cores of the Perseus Molecular Cloud focusing on c-C$_3$H$_2$ and its deuterated isotopologues.\

This study is structured as follows: the observations are presented in Section \ref{observations}, the description of the methodology used to reduce and fit the data are presented in Section \ref{analysis}, the results are presented in Section \ref{results}, these results are interpreted in Section \ref{discussion}, and we summarize the conclusion in Section \ref{conclusions}. Moreover, supplementary information is collected in the Appendix.\

\section{Observations}\label{observations}

\begin{figure*}[h]
\centering
\includegraphics[width=\textwidth]{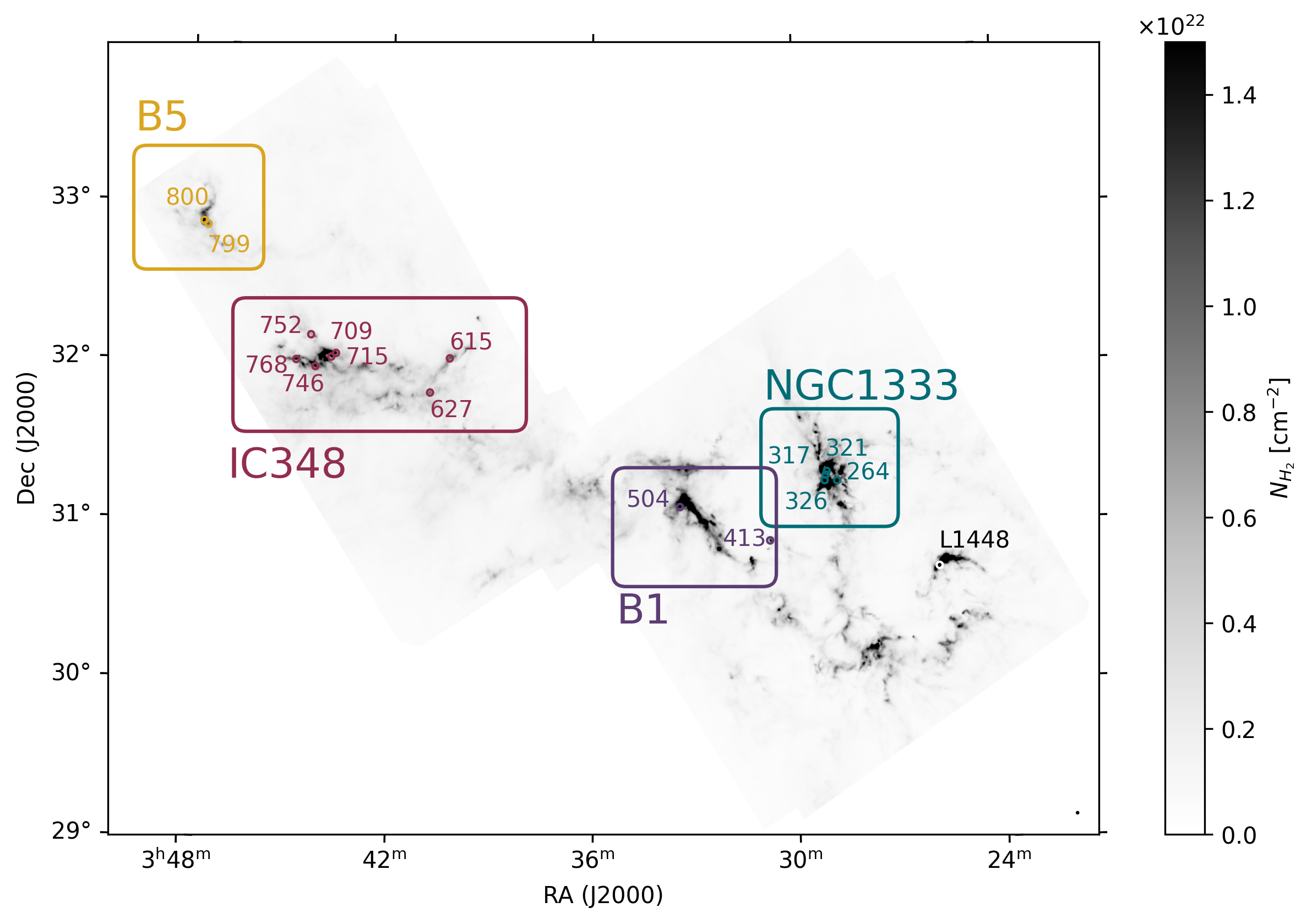}
\caption{Molecular hydrogen column density ($N_{\rm H_2}$) map of the Perseus Molecular Cloud in gray scale \citep{pezzuto:12,sadavoy:12,sadavoy:14}. The starless and prestellar cores studied in this study are indicated by coloured circles corresponding to the regions within the molecular cloud where they are located. NGC1333 is plotted in teal, B1 in purple, IC348 in red and B5 in yellow.}
\label{map}
\end{figure*}

Single pointing observations of c-C$_{3}$H$_{2}$, c-C$_{3}$HD and c-C$_{3}$D$_{2}$ towards fifteen Perseus cores were taken by two different telescopes: the Yebes Observatory 40m telescope and the Arizona Radio Observatory (ARO) 12m telescope. Additionally, supplemental data from the Institut de Radioastronomie Millimétrique 30 meter telescope (IRAM 30m) for one additional core is used, bringing the total number of cores to sixteen (Figure \ref{map}, Table \ref{source}). All of the transitions targeted are presented in Table \ref{obsdat}.  \

The data taken by the Yebes 40m telescope was first presented in \citealt{scibelli:24}. These observations were taken between 2022 and 2023 corresponding to the projects with ID 22A022 and 23A025 (PI: Scibelli). The fifteen Perseus cores targeted were selected on the basis of the detection of acetaldehyde (CH$_{3}$CHO) to search for other complex organic molecules (COMs). These cores were observed with the Q-band wide-band receiver \citep{tercero:21} with a bandwidth of 18.5 GHz in the 31.5 – 50 GHz frequency range. The resolution of the observations is of 38.0 kHz (0.38 km s$^{-1}$ – 0.23 km s$^{-1}$). The beam ranges between 36$^{\prime\prime}$ and 56$^{\prime\prime}$ across the full frequency range. The data was reduced and converted to main beam temperature ($T_{\rm MB}$) using the publicly available Python-based scripts\footnote{\url{https://github.com/andresmegias/gildas-class-pipeline/}} developed by \citealt{megias:23}, which calls on the Continuum and Line Analysis Single-dish Software (\textsc{class}), an application from the \textsc{gildas}\footnote{\url{https://www.iram.fr/IRAMFR/GILDAS/}} software \citep{pety:05}. For further information on the data and on the reduction process refer to \citealt{scibelli:24}.\

Moreover, to add additional transitions, we observed c-C$_{3}$H$_{2}$ and c-C$_{3}$HD transitions using the ARO 12-m Radio Telescope on Kitt Peak (PI: Steffes). Observations occurred over 46 shifts from February 21, 2025 through June 15, 2025 using absolute position switching for 30 seconds on and off source for 5 minute scans. Using the multi-window mode on AROWS, we observed two transitions of para-c-C$_{3}$H$_{2}$, these being the 2$_{2, 0}$ - 1$_{1, 1}$ (150.44 GHz) and the 4$_{0, 4}$- 3$_{1, 3}$ (150.820665 GHz) transitions. Then, we observed two transitions of ortho-c-C$_{3}$H$_{2}$, these being the 4$_{1, 4}$ - 3$_{0, 3}$ (150.85 GHz) and the 3$_{1, 2}$ - 2$_{2, 1}$ (145.09 GHz) transitions. Finally, we observed to two transitions of c-C$_{3}$HD. These were the 4$_{1, 4}$ - 3$_{0, 3}$ (136.370909 GHz) and 2$_{2, 0}$ - 1$_{1, 1}$ (137.454464 GHz) transitions. After applying Hanning smoothing, the velocity resolution ranged over 0.08 - 0.09 km s$^{-1}$ with a frequency resolution of 39.06 kHz. These observations had a beam size ranging from 41$^{\prime\prime}$ to 46$^{\prime\prime}$.\

The length of the ARO 12m observations was determined by the amount of time required for each to achieve a signal-to-noise ratio (SNR) $\geq$ 5. Because of the use of multi-window mode, this resulted in many observations achieving a SNR > 5 as we stayed on source in a given setup until the weakest transition achieved a SNR of 5. Several sources were simply too weak to achieve this sensitivity within the given time. Linear functions were then fit to all of the velocity channels not within the 5$\sigma$ boundaries of the Gaussian. All of these observations were summed, weighted by 1/$\sigma^{2}$ (with $\sigma$ being the RMS noise level) and fit with a Gaussian using \textsc{class}. These were then subtracted from the summed spectra to baseline them. The average beam efficiency for the 2mm frequency range of 81\% was applied.\

Additionally, we are including one c-C$_{3}$HD and one c-C$_{3}$D$_{2}$ transition observed towards another prestellar core in the Perseus Molecular Cloud, L1448.  These observations were carried out with the Institut de Radioastronomie Millimétrique 30 meter telescope (IRAM 30m) within the project with ID  IRAM 071-23 (PI: Fuente, \cite{kulterer:25}). The observations were carried out on the 2nd and 3rd of May 2024 in excellent weather (PWV = 0–4.6 mm, $\tau$ $<$ 0.1) and at a system temperature ($T_{\rm sys}$) of 60–130 K. EMIR with the FTS backend and a spectral resolution of 49 kHz was used. The data were taken in frequency-switching mode with a frequency throw of 3.9 MHz. Of the four original spectral windows, those centred on 93.6 and 106.1 GHz, with rms of 2.13 and 2.20 mK, were used. The half power beam width (HPBW) of c-C$_{3}$HD 3$_{03}$ - 2$_{02}$ is 23" and the one for the c-C$_{3}$D$_{2}$ 3$_{03}$ - 2$_{12}$ transition is 26".\

\begin{table*}[ht]
\begin{center}
\caption{Starless and prestellar cores in the Perseus Molecular Cloud targeted in this study. }

\begin{tabular}{ lccccccc } 
\hline\hline
Region & Core $\#$ $^{a}$ & RA & DEC &  Distance$^{b}$ & $N_{H_{2}}$ & $n_{H_{2}}$  & $T_{\rm kin}$  \\
 & (\textit{Herschel}) & (J2000) & (J2000) & (pc) & ($\times$ 10$^{22}$ cm$^{-2}$) & ($\times$ 10$^{5}$ cm$^{-3}$) & (K) \\
\hline\hline
\textcolor{NGC1333}{NGC1333} & 264 & 03:28:47.14 & +31:15:11.4 & 294 & 2.06 & 0.75 & 10.8  \\
& 317 & 03:29:04.93 & +31:18:44.4 & 294 & 2.63 & 0.96 & 13.6  \\
& 321 & 03:29:07.17 & +31:17:22.1 & 294 & 3.65 & 1.33 & 12.6  \\
& 326 & 03:29:08.97 & +31:15:17.2 & 294 & 7.88 & 2.87 & 12.3  \\
\hline
\textcolor{B1}{B1} & 413 & 03:30:46.74 & +30:52:44.8 & 297 & 1.17 & 0.42 & 10.5  \\
& 504 & 03:33:25.31 & +31:05:37.5 & 297 &1.80 & 0.65 & 9.7  \\
\hline
\textcolor{IC348}{IC348} & 615 & 03:40:14.92 & +32:01:40.8 & 314 & 1.03 & 0.35 & 10.3  \\
& 627 & 03:40:49.53 & +31:48:40.5 & 314 & 0.93 & 0.32 & 12.4  \\
& 709 & 03:43:38.06 & +32:03:07.4 & 314 & 1.48 & 0.51 & 13.5  \\
& 715 & 03:43:46.34 & +32:01:43.5 & 314 & 1.45 & 0.49 & 14.8  \\
& 746 & 03:44:14.38 & +31:58:00.7 & 314 & 1.42 & 0.49 & 10.6  \\
& 752 & 03:44:23.10 & +32:10:01.0 & 314 & 0.63 & 0.22 & 14.7  \\
& 768 & 03:44:48.83 & +32:00:31.6 & 314 & 1.43 & 0.49 & 10.8  \\
\hline
\textcolor{B5}{B5} & 799 & 03:47:31.31 & +32:50:56.9 & 325 & 1.01 & 0.33 & 10.4  \\
& 800 & 03:47:38.97 & +32:52:16.6 & 325 & 1.93 & 0.64 & 11.7  \\
 \hline
& L1448$^{c}$ & 03:25:49.00 & +30:42:24.6 & 285 & 10.6 & 0.57 & 15.1 \\
\hline
\label{source}
\end{tabular}

\tablefoot{ $^{a}$ The cores are numbered in agreement with \cite{pezzuto:21}. $^{b}$ The distances of the cores are taken from \cite{pezzuto:21}. $^{c}$ The physical properties of L1448 were obtained through private communication related to the work of \cite{rodriguez:21}. The different subregions of the Perseus Molecular Cloud are colour coded to ease posterior analysis.}
\end{center}
\end{table*}

\begin{table*}[ht]
\begin{center}
\caption{Observed Molecular Transitions}

\begin{tabular}{ lccccc } 
\hline\hline
Molecule & Transition &  & Frequency &  $E_\mathrm{up}$ & $A_{ij}$  \\
 & & (o/p/-) & (MHz) & (K) & ($\times$ 10$^{-6}$ s$^{-1}$)  \\
\hline\hline
\textbf{Yebes 40m} & & & & &  \\
c-C$_{3}$H$_{2}$ & 3$_{21}$ - 3$_{12}$ & o & 44104.7769 & 30.85 & 3.20   \\
 & 2$_{11}$ - 2$_{02}$ & p & 46755.6100 & 8.67 & 2.68   \\
c-C$_{3}$HD & 2$_{11}$ - 2$_{02}$ & - & 38224.4415 & 7.56 & 1.75   \\
 & 1$_{01}$ - 0$_{00}$ & - & 42064.1497 & 2.02 & 0.42   \\
 & 1$_{11}$ - 0$_{00}$ & - & 49615.8607 & 2.38 & 4.38   \\
c-C$_{3}$D$_{2}$ & 2$_{11}$ - 2$_{02}$ & o & 34477.1066 & 6.84 & 1.52  \\
& 1$_{11}$ - 0$_{00}$ & o & 45358.8400 & 2.18 & 3.87   \\
\hline
\textbf{ARO 12m} & & & & & \\
c-C$_{3}$H$_{2}$ & 2$_{20}$ - 1$_{11}$ & p & 150436.5547 & 9.71 & 53.57    \\
& 4$_{04}$ - 3$_{13}$ & p & 150820.6650 & 19.31 & 163.64    \\
& 3$_{12}$ - 2$_{21}$ & o & 145089.6055 & 16.05 & 135.62   \\
& 4$_{14}$ - 3$_{03}$ & o & 150851.9080 & 19.31 & 163.77    \\
c-C$_{3}$HD & 4$_{14}$ - 3$_{03}$ & - & 136370.9100 & 17.39 & 137.24    \\
& 2$_{20}$ - 1$_{11}$ & - & 137454.4640 & 8.98 & 102.75  \\
\hline
\textbf{IRAM 30m} & & & & &   \\
c-C$_{3}$HD & 3$_{03}$ - 2$_{02}$ & - & 106811.0901 & 10.85 & 7.87    \\
c-C$_{3}$D$_{2}$ & 3$_{03}$ - 2$_{12}$ & p &  94371.3538 & 9.85 & 33.73 \\
 \hline
\label{obsdat}
\end{tabular}

\tablefoot{ The spectral information for the molecular transitions including the frequency, the upper energy, ($E_{\rm up}$), and the Einstein coefficient, ($A_{ij}$), was taken from The Cologne Database for Molecular Spectroscopy (CDMS)\footnote{\url{https://cdms.astro.uni-koeln.de}} (c-C$_{3}$H$_{2}$: \citealt{bogey:86, vrtilek:87, lovas:92, spezzano:12}, c-C$_{3}$HD: \citealt{bogey:86, spezzano:12}, c-C$_{3}$D$_{2}$: \citealt{spezzano:12}). For each transition an o, p or - indicates whether the transition is ortho, para or non-applicable.}
\end{center}
\end{table*}

\section{Analysis} \label{analysis}
The observations were fit with the Python "pyspeckit" package \citep{ginsburg:11, ginsburg:22}. For three transitions, c-C$_{3}$H$_{2}$ 3$_{2,1}$ - 3$_{1,2}$, c-C$_{3}$D$_{2}$ 1$_{1,1}$ - 0$_{0,0}$ for the core Per799, c-C$_{3}$D$_{2}$ 1$_{1,1}$ - 0$_{0,0}$ for the core Per627, the fit resulted in anomalously large errors for the $V_{LSR}$ and the full width half maximum (FWHM). The limited spectral resolution of the Yebes 40m observations, which ranges between 0.23 and 0.38 km s$^{-1}$ compared to the 0.08 - 0.09 km s$^{-1}$ of the ARO 12m and 0.14-0.16 km s$^{-1}$ of the IRAM 30m observations, does not allow to resolve these lines, which appear narrower than the rest. For these three transitions the area under the line is directly derived from the spectra with a Python routine, and then by assuming a fixed line width, the peak temperatures are calculated and printed in Tables \ref{tcc3h2} and \ref{tcc3d2} alongside an asterisk (*). A transition is considered detected if its peak temperature is $\geq$3$\sigma$. The c-C$_{3}$H$_{2}$ and c-C$_{3}$HD isotopologues are considered detected towards a core if multiple transitions were detected. In the case of c-C$_{3}$D$_{2}$, we consider that if the main and D-isotopologues have been detected towards a specific core, and we have at least one detected c-C$_{3}$D$_{2}$ transition, this isotopologue is also detected.\

For the derivation of the column density, it should be noted that both c-C$_{3}$H$_{2}$ and c-C$_{3}$D$_{2}$ have ortho and para states. Because the interconversion timescales within these states are long, ortho and para states can be treated as separate molecules. Thus, the total column density of c-C$_{3}$H$_{2}$ is the column density of ortho-c-C$_{3}$H$_{2}$ plus the column density of para-c-C$_{3}$H$_{2}$. The same applies to c-C$_{3}$D$_{2}$. For more information, refer to Section \ref{otprat}.\

The column densities for the different isotopologues were calculated using the RADEX code \citep{vandertak:07b}. RADEX is a non-LTE radiative transfer code that requires physical information about the source, volume density ($n_{\rm H_2}$) and kinetic temperature ($T_{\rm kin}$), as well as spectral and collisional rate information of the molecule to model its transitions.\

The input $n_{\rm H_2}$, and $T_{\rm kin}$ used to model the line intensity, and subsequently constrain column densities, (see Table \ref{source}) are based on the median values for \textit{Herschel} maps \citep{pezzuto:21} within the ARO 12m 62'' beam. The data analysed in this study has different beam sizes owing to the use of different telescopes and different targeted frequencies. We take the derived volume densities listed in \cite{scibelli:24} as their value does not significantly change when considering the different beam sizes in this study (see Section \ref{columndensity} in the Appendix). Additionally, we compute the $n_{\rm H_2}$ value for L1448 also with a \textit{Herschel} map using a 62'' beam, for comparison purposes. The $T_{\rm kin}$ values for different cores are constrained from NH$_{3}$ observations, except for Per615 where the \textit{Herschel} $T_{\rm dust}$ is reported (see \cite{scibelli:24} for details). The $T_{\rm kin}$ for L1448 is derived from far infrared and sub-millimetre observations \citep{zari:16, rodriguez:21}.\

For the three transitions where the pyspeckit fitting was not possible, the area extracted under the line was used alongside an assumed linewidth (0.5 km s$^{-1}$) to derive the line intensity. A 10\% uncertainty was assumed for both $n_{\rm H_2}$ and $T_{\rm kin}$ (see \cite{scibelli:24}). The column density values do not vary significantly for a sensible range of $n_{\rm H_2}$ and $T_{\rm kin}$ (for more information see Appendix \ref{columndensity}). The column density for each molecule (ortho-c-C$_{3}$H$_{2}$, para-c-C$_{3}$H$_{2}$, c-C$_{3}$HD, ortho-c-C$_{3}$D$_{2}$ and para-c-C$_{3}$D$_{2}$) was adjusted minimizing the $\chi ^{2}$ value. Uncertainties were estimated using a Monte Carlo approach in which the observed line parameters, $n_{\rm H_2}$ and $T_{\rm kin}$ were randomly perturbed within their respective uncertainties, and the RADEX fit was repeated for each trial. Lastly, the total column densities were calculated for c-C$_{3}$H$_{2}$ by summing the ortho and para column densities. In the case of c-C$_{3}$D$_{2}$ there are no ortho and para transitions observed towards the same source simultaneously, so a statistical ortho-to-para ratio value of 2 is used to compute the total column densities (see Section \ref{otprat}).   \

Finally, to explore possible correlations between the physical parameters of the cores and the derived column densities, otp ratios, D/H and D$_{2}$/D ratios we employ the Pearson correlation coefficient ($r$), which measures statistically the linear relationship between two variables.

\section{Results} \label{results}

\begin{figure*}[h]
\centering
\includegraphics[width=\textwidth]{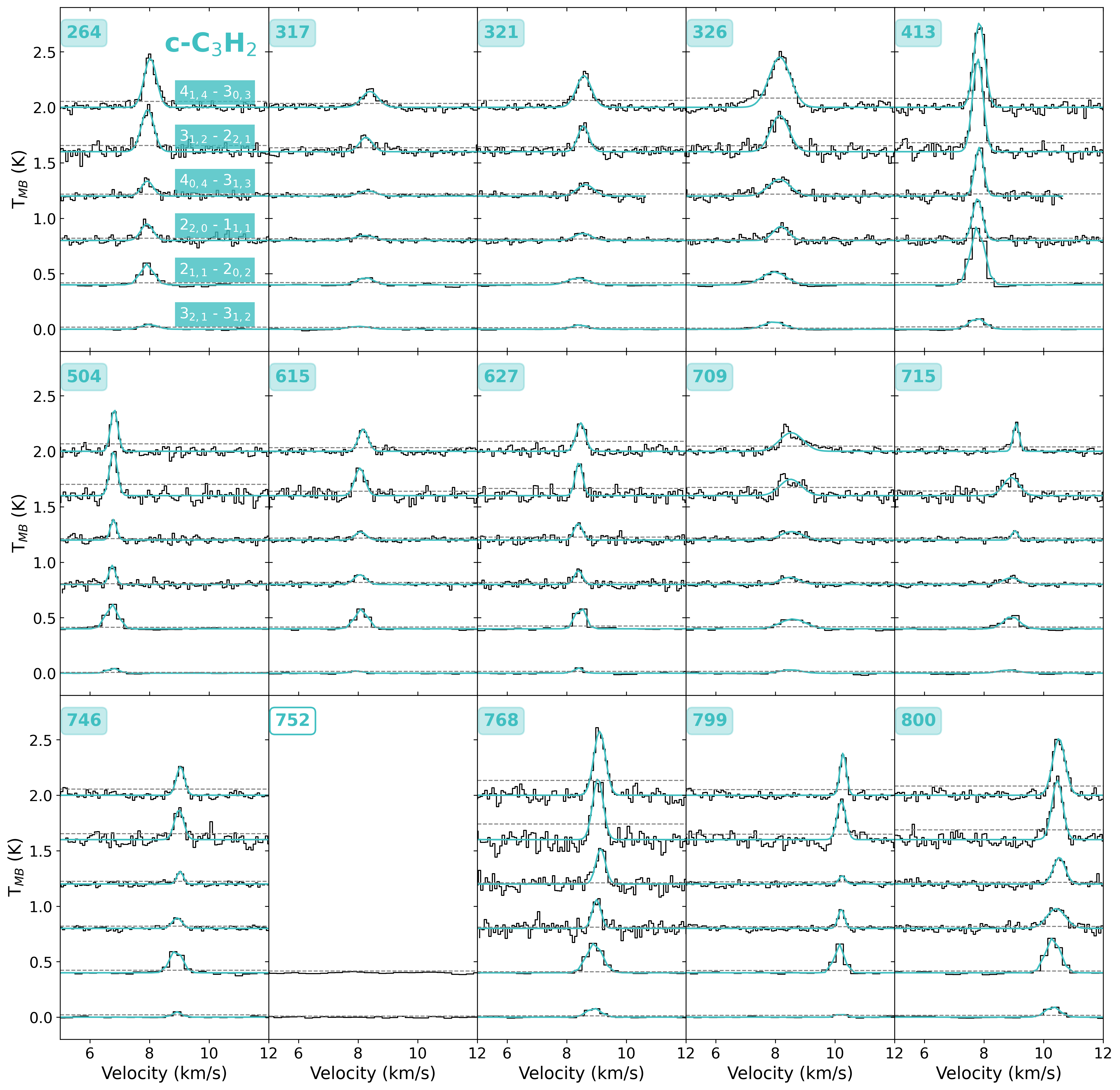}
\caption{Observed spectra of the c-C$_{3}$H$_{2}$ transitions organised by core number. All of the transitions for the same core are plotted in the same subplot with vertical offsets for clarity. The Gaussian fit from pyspeckit are overplotted with a blue line. The spectra without a Gaussian fit indicate a non-detection. Core number labels with a colour background indicate c-C$_{3}$H$_{2}$ has been detected towards the source. Lastly, horizontal gray dashed lines indicates 3$\sigma$ levels.}
\label{occ3h2}
\end{figure*}

\begin{figure*}[h]
\centering
\includegraphics[width=\textwidth]{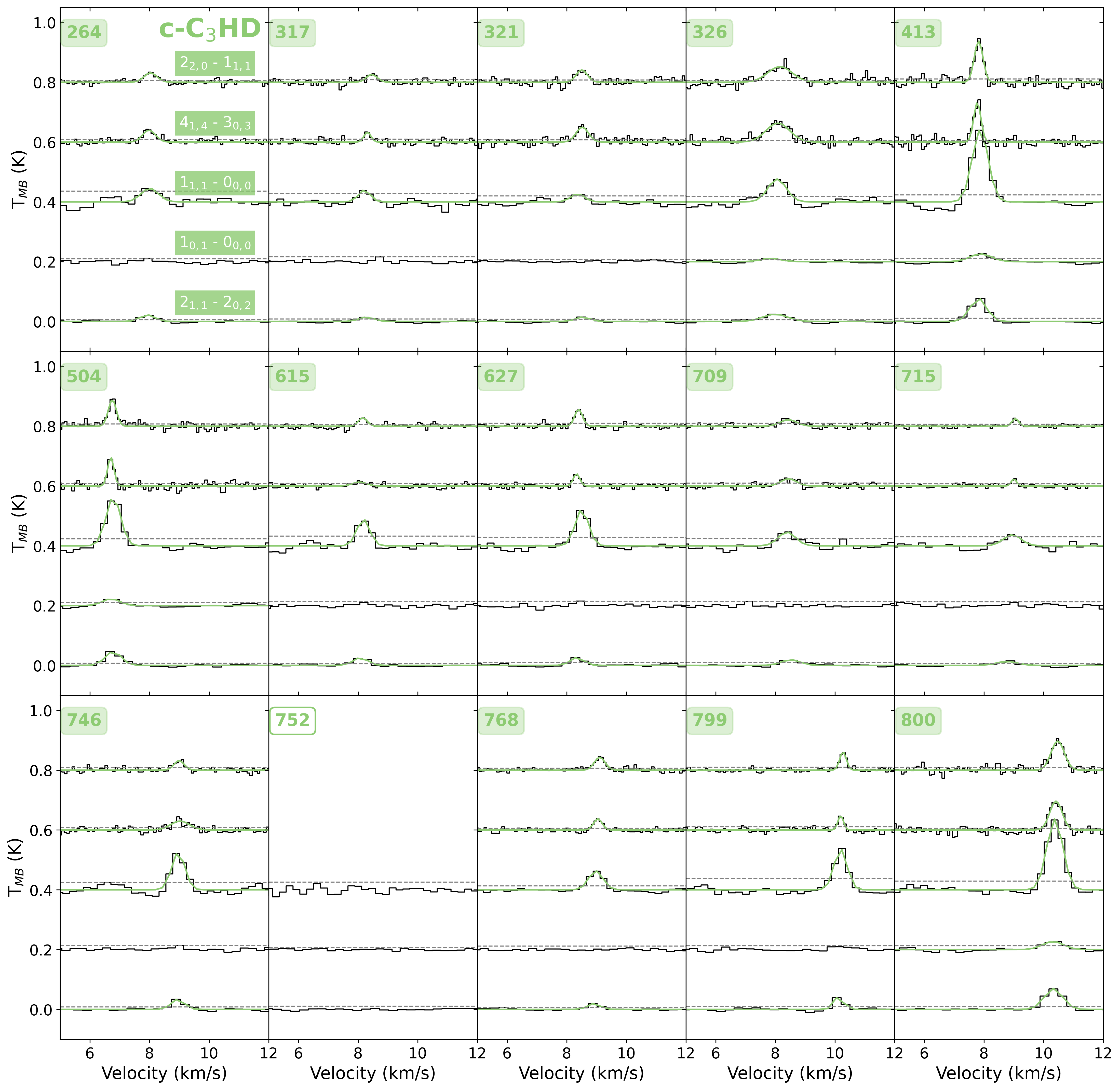}
\caption{Same as Figure \ref{occ3h2} for c-C$_{3}$HD. Here the Gaussian fits and labels are plotted with the colour green. }
\label{occ3hd}
\end{figure*}

\begin{figure*}[h]
\centering
\includegraphics[width=\textwidth]{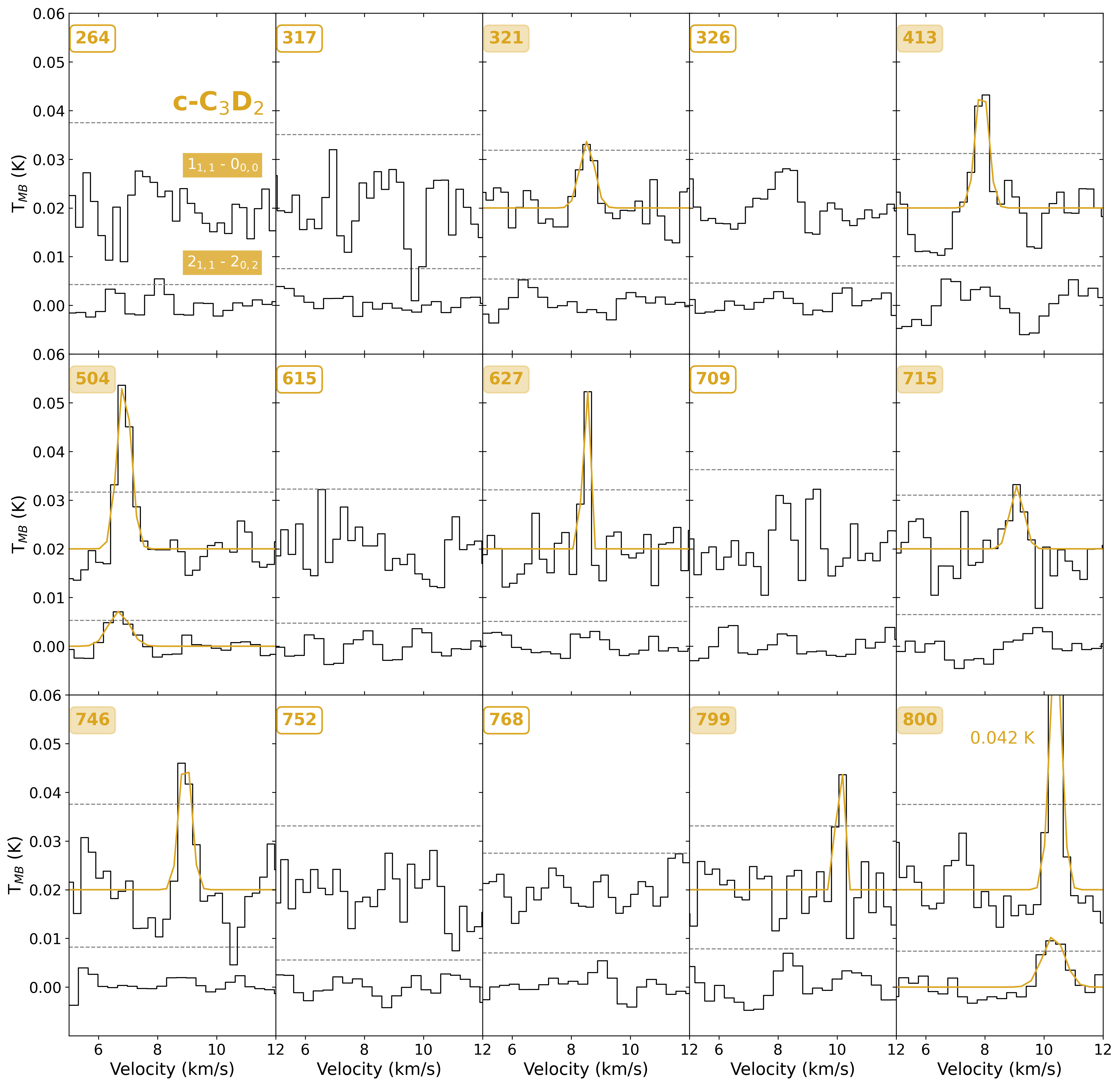}
\caption{Same as Figure \ref{occ3h2} for c-C$_{3}$D$_{2}$. Here the Gaussian fits and labels are plotted with the colour orange. }
\label{occ3d2}
\end{figure*}

\begin{figure*}[h]
\centering
\includegraphics[width=\textwidth]{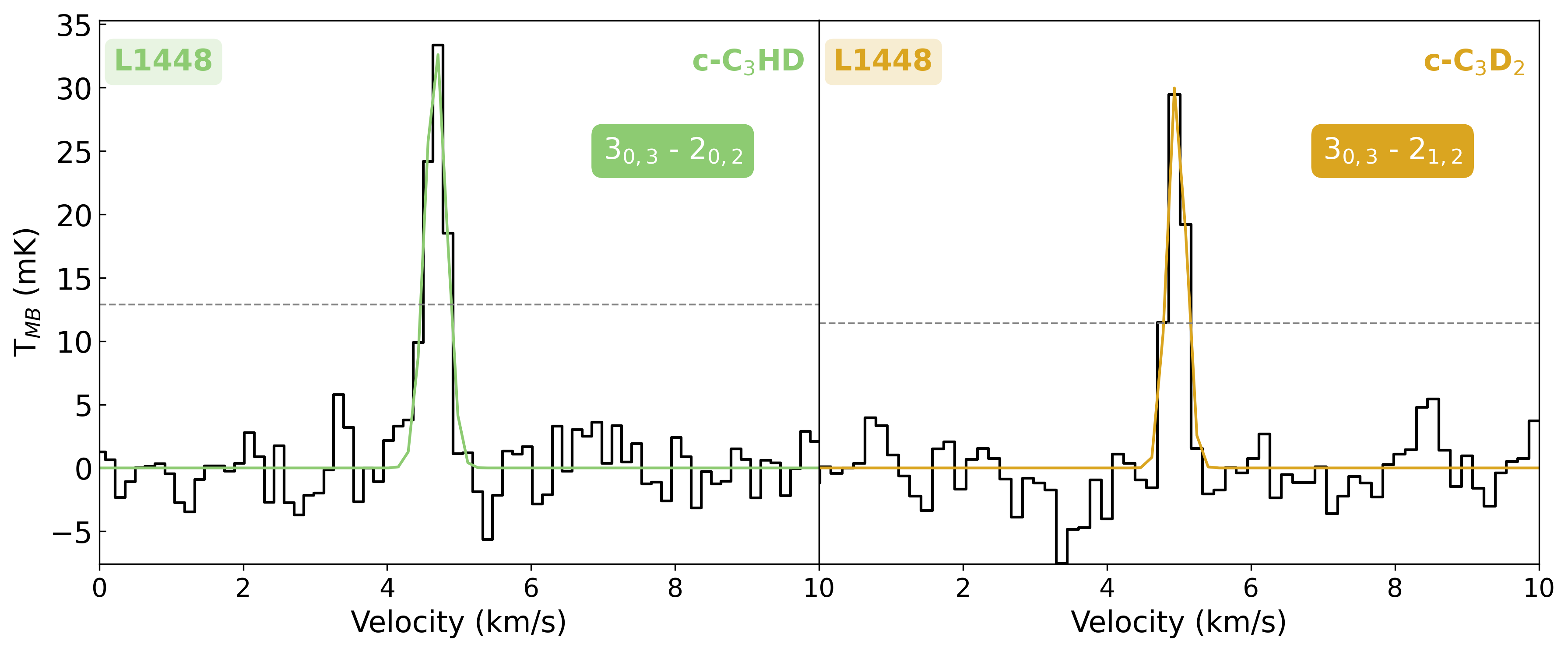}
\caption{Observed spectra and the Gaussian fit from pyspeckit of c-C$_{3}$HD, in green, and c-C$_{3}$D$_{2}$, in orange. Horizontal gray dashed lines indicates 3$\sigma$ levels.}
\label{ol1448}
\end{figure*}

\begin{figure*}[h]
\centering
\includegraphics[width=\textwidth]{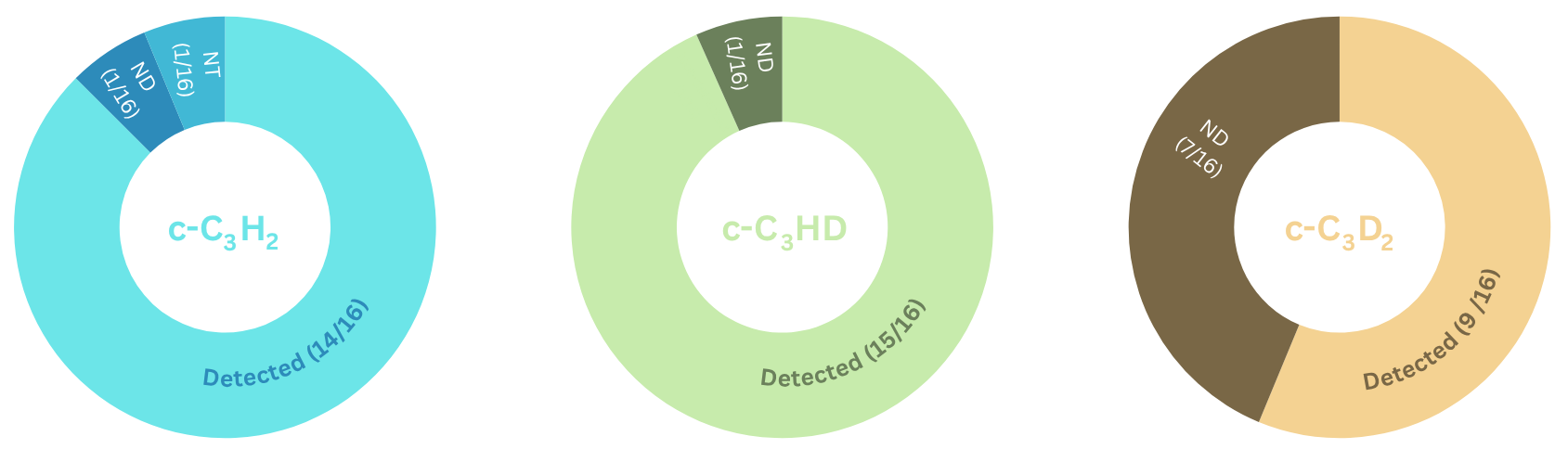}
\caption{Detection statistics of c-C$_{3}$H$_{2}$, in blue, c-C$_{3}$HD, in green, and c-C$_{3}$D$_{2}$, in orange. The pie chart shows three categories: Detected, Not-detected (ND), and Not-targeted (NT). c-C$_{3}$H$_{2}$ and c-C$_{3}$HD were not targeted towards one of the cores, while C$_{3}$D$_{2}$ was targeted towards all of them.  }
\label{pie}
\end{figure*}

The parameters resulting from the Gaussian fitting of the observed c-C$_{3}$H$_{2}$, c-C$_{3}$HD and c-C$_{3}$D$_{2}$ towards all of the cores ($T_{\rm MB}$, $V_{LSR}$, FWHM and rms ) can be found in Appendix \ref{obstran}. The fitting of the transitions towards the Perseus cores can be seen in Figures \ref{occ3h2}, \ref{occ3hd} and \ref{occ3d2}, except for core L1448 which is plotted in Figure \ref{ol1448}. The number of detected transitions for each of the isotopologues towards each source is summarised in Table \ref{dettrans}, and the total detection statistics, including detected, non-detected and non-targeted, are shown with pie charts in Figure \ref{pie}.\

The column densities for each of the isotopologues towards each of the cores, calculated following the method described in Section \ref{analysis}, are summarised in Table \ref{coldenst}.\

The results for each of the isotopologues, c-C$_{3}$H$_{2}$, c-C$_{3}$HD and c-C$_{3}$D$_{2}$ are discussed in Sections \ref{cc3h2}, \ref{cc3hd}, \ref{cc3d2}, respectively. Moreover, the c-C$_{3}$H$_{2}$ ortho-to-para ratio and its  deuteration ratios are presented in Subsections \ref{otprat} and \ref{deutrat}.\

The plots related to c-C$_{3}$H$_{2}$, c-C$_{3}$HD and c-C$_{3}$D$_{2}$ are plotted consistently through this study in blue, green and orange, respectively. \

\begin{table}[ht]
\begin{center}
\caption{Number of detected transitions per isotopologue and core.}

\begin{tabular}{ lccc } 
\hline\hline
Core Number & c-C$_{3}$H$_{2}$ & c-C$_{3}$HD & c-C$_{3}$D$_{2}$  \\
\hline
264 & 6 & 4 & 0  \\
317 & 6 & 4 & 0  \\
321 & 6 & 4 & 1  \\
326 & 6 & 5 & 0  \\
413 & 6 & 5 & 1  \\
504 & 6 & 5 & 2  \\
615 & 6 & 4 & 0  \\
627 & 6 & 4 & 1  \\
709 & 6 & 4 & 0  \\
715 & 6 & 4 & 1  \\
746 & 6 & 4 & 1  \\
752 & 0 & 0 & 0  \\
768 & 6 & 4 & 0  \\
799 & 6 & 4 & 1  \\
800 & 6 & 5 & 2  \\
L1448 & -$^{*}$ & 1 & 1  \\
\hline
\label{dettrans}
\end{tabular}

\tablefoot{A transition is considered to be detected when its peak temperature $\geq$3$\sigma$. $^{*}$ L1448 was not targeted for c-C$_{3}$H$_{2}$.}
\end{center}
\end{table}

\subsection{c-C$_{3}$H$_{2}$} \label{cc3h2}

For c-C$_{3}$H$_{2}$ a total of six lines were targeted: two at lower frequencies with the Yebes 40m telescope: 3$_{21}$ - 3$_{12}$ (o) and 2$_{11}$ - 2$_{02}$ (p); and four at higher frequencies with the ARO 12m telescope:  2$_{20}$ - 1$_{11}$  (p), 4$_{04}$ - 3$_{13}$ (p), 3$_{12}$ - 2$_{21}$ (o) and 4$_{14}$ - 3$_{03}$ (o). The additional ARO 12m observations were done a posteriori to have multiple ortho and para transitions with different $E_{up}$ to be able to better constrain the excitation temperature and column density of c-C$_{3}$H$_{2}$. This molecule was not targeted towards L1448.\

Out of the 15 cores where our data covers c-C$_{3}$H$_{2}$, we detected the molecule towards 14 cores ($\sim$ 93\%). The SNR calculated from the line peak temperature, ranges from 3 to 86 for the detected transitions. The only core where c-C$_{3}$H$_{2}$ was not detected is Per752.\

Note that some lines observed with the ARO 12m telescope towards Per709 have profiles that deviate from a Gaussian. This line profile is resolved for some of the transitions observed with the ARO 12m telescope, but not for the transitions observed with the Yebes 40m telescope, owing to their different spectral resolutions. The origin of this double-peaked line profiles are unknown. Because this profile is only marginally resolved in a few transitions, we do not attempt to model it. We note that this introduces some additional uncertainty in the column densities derived towards Per709.  \

The 3$_{2,1}$-3$_{1,2}$ ortho transition could not be modelled with RADEX for core Per413. The c-C$_{3}$H$_{2}$ 3$_{2,1}$-3$_{1,2}$ transition has quite a higher upper energy (E$_{up}$ $\sim$30 K) compared to the other ortho transitions. Per413 is the core where c-C$_{3}$H$_{2}$ is the most abundant (Figure \ref{abun} in Appendix \ref{abundance}). Thus, in Per413 this transition may be tracing warmer gas from the envelope at different $n_{\rm H_2}$ and $T_{\rm kin}$ and thus could not be fitted alongside the other ortho transitions. This transition has been excluded from the column density calculation for Per413.\

The column density of c-C$_{3}$H$_{2}$ was calculated using the method described in Section \ref{analysis}. The column densities for the ortho- and para-c-C$_{3}$H$_{2}$ were computed separately, as they have different collisional rate coefficients. The main isotopologue rate coefficients are given in \cite{khalifa:19}, while the ones for the deuterated isotopologues are given in Ben Khalifa et al. (in prep.). Then, these obtained values are summed to give the total-c-C$_{3}$H$_{2}$ column density which we present in Table \ref{coldenst}. For the core Per752, as c-C$_{3}$H$_{2}$ was not detected, we give an upper limit column density. This core was exclusively observed with the Yebes 40m data, and thus, the upper limit column density is estimated with just the 3$_{2,1}$ - 3$_{1,2}$ and 2$_{1,1}$ - 2$_{0,2}$ transitions. A line width of 0.5 km s$^{-1}$ was assumed. The total c-C$_{3}$H$_{2}$ column density values found range from 5.0 $\pm$ 1.0$\times 10^{12}$ cm$^{-2}$ (Per317) to 8.1 $\pm$ 1.8$\times 10^{13}$ cm$^{-2}$ (Per413), with a weighted average of 8.6 $\pm$ 0.5$\times 10^{12}$ cm$^{-2}$ (not taking into account the upper limit of Per752). \

The ortho-c-C$_{3}$H$_{2}$ excitation temperatures ($T_{\rm ex}$) returned by RADEX are very similar amongst transitions and cores with a median of 3.7 K. The situation is similar for para-c-C$_{3}$H$_{2}$, where the $T_{\rm ex}$ has a median of 3.5 K. The optical depth ($\tau$) of c-C$_{3}$H$_{2}$ changes for the different transitions. For 3$_{2,1}$ - 3$_{1,2}$, $\tau$ has a median of 0.08, for 2$_{1,1}$ - 2$_{0,2}$ the median $\tau$ is 0.14, for 2$_{2,0}$ - 1$_{1,1}$ is 0.73, for 4$_{0,4}$ - 3$_{1,3}$ is 0.32, for 3$_{1,2}$ - 2$_{2,1}$ the median is 0.83 and for 4$_{1,4}$ - 3$_{0,3}$ the median is 1.27.\ 

\subsubsection{c-C$_{3}$H$_{2}$ ortho-to-para ratio} \label{otprat}

Molecules with symmetry that have two equivalent protons (or H atoms) exhibit two different type of states: ortho and para. These different states arise from the orientation of the two protons' spin directions. Ortho corresponds to nuclear wavefunctions that have a total spin $I_{\rm tot}$ = 1 and are symmetric to interchange of the protons. Para nuclear wavefunctions have  $I_{\rm tot}$ = 0 and are antisymmetric to interchange of the protons. The interconversion between ortho and para is low, due to the weakness of the nuclear magnetic interaction, which makes it possible to treat ortho and para states as two different molecules.\

The c-C$_{3}$H$_{2}$ molecule has two equivalent protons, which consequentially, results in a separation of ortho and para states. The spin of the hydrogen atom is $\pm$ 1/2. If we take into account the relation between the spin of the atom and the total spin of the molecule, $I_{\rm tot}$ = |I$_{1}$ - I$_{2}$|, ... , I$_{1}$ + I$_{2}$, where I$_{1,2}$ are the nuclear spin of each atom, the total nuclear spin of c-C$_{3}$H$_{2}$ can be either 0 or 1. Knowing that the number of individual states is given by 2$I_{\rm tot}$ + 1, then for $I_{\rm tot}$ = 0 we get 1 state and for $I_{\rm tot}$ = 1 we get 3 states. In this case $I_{\rm tot}$ = 0 correspond to the para states and $I_{\rm tot}$ = 1 correspond to the ortho states. Then, statistically, the ortho-to-para ratio of c-C$_{3}$H$_{2}$ is 3. Due to its symmetry, for c-C$_{3}$H$_{2}$, the states that have quantum numbers $K_a$ and $K_c$ such that $K_a$ + $K_c$ = odd value will be ortho and the states that have $K_a$ + $K_c$ = even will be para. This is why for example the 3$_{21}$ - 3$_{12}$ transition (see Table \ref{obsdat}) is an ortho transition. Nevertheless, the ortho-to-para ratio can be non-statistical due to the physical conditions of the environment. For example, the H$_{2}$ otp ratio is seen to be lower than 3 in prestellar cores ($<$ 0.01; \citealp{pagani:09, dislaire:12}) due to the low temperatures present. The protons in the c-C$_{3}$HD molecule are not equivalent and thus, do not present ortho and para substates. On the other hand the protons in c-C$_{3}$D$_{2}$ are equivalent and this molecule will show separate ortho and para states. The spin in the case of deuterium is 1. Following the same reasoning as for c-C$_{3}$H$_{2}$, the $I_{\rm tot}$ = 0, 1, 2, with a number of states equal to 1, 3 and 5, respectively. In the case of c-C$_{3}$D$_{2}$ $I_{\rm tot}$ = 0 and 2, accounting for 6 symmetric nuclear wavefunctions, are ortho, and $I_{\rm tot}$ = 1, accounting for 3 anti-symmetric wavefunctions, are para. Thus, the statistical ratio of c-C$_{3}$D$_{2}$ is 2. Due to the symmetry of c-C$_{3}$D$_{2}$, the states that have $K_a$ + $K_c$ = odd value will be para and the ones that have that have $K_a$ + $K_c$ = even will be ortho. Notice that the $K_a$ + $K_c$ rule is opposite for the c-C$_{3}$H$_{2}$ and c-C$_{3}$D$_{2}$ molecules. As ortho- and para-c-C$_{3}$D$_{2}$ were not detected both towards the same cores, its ortho-to-para ratio could not be studied.\

\begin{figure*}[h]
\centering
\includegraphics[width=\textwidth]{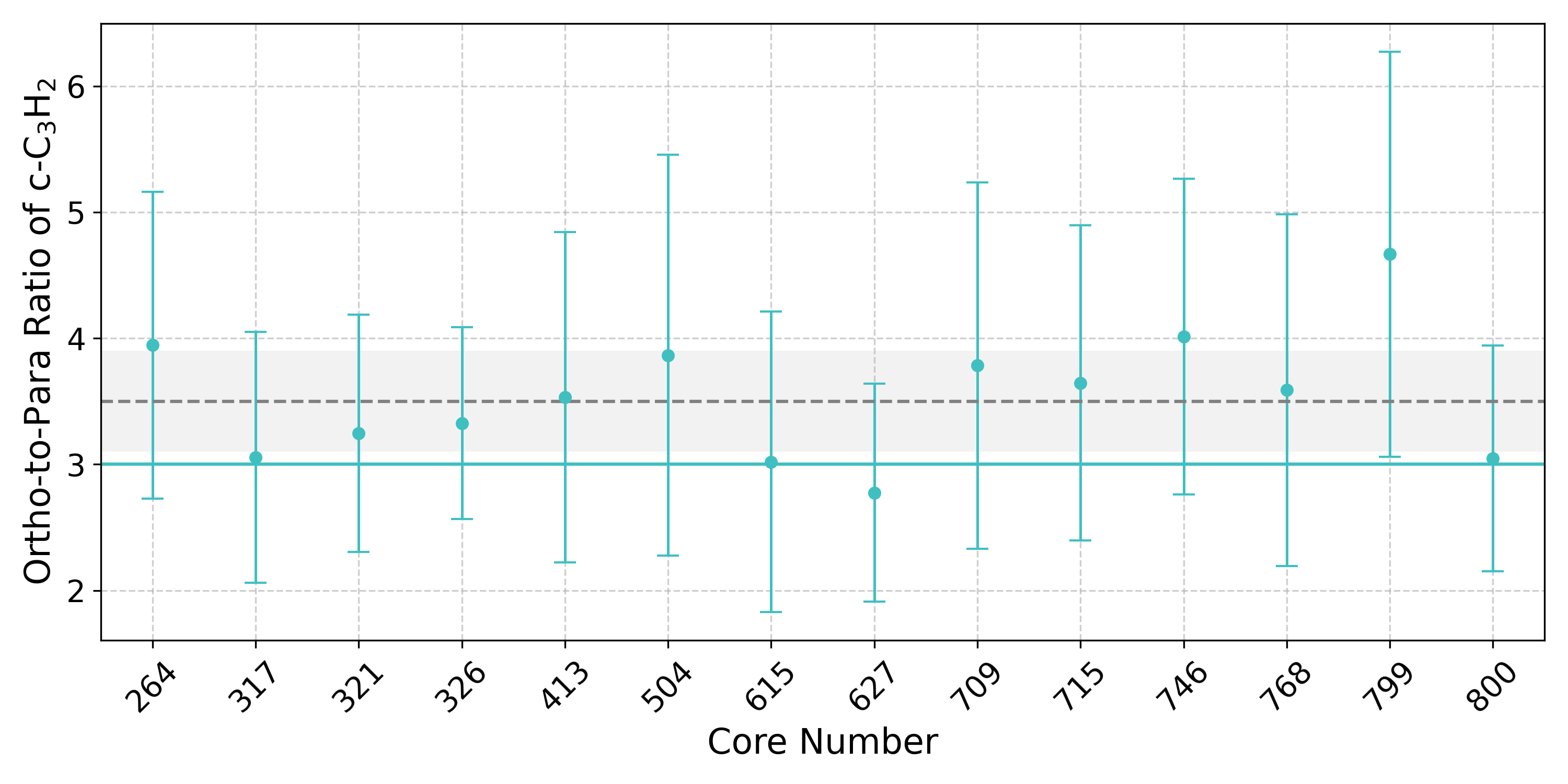}
\caption{c-C$_{3}$H$_{2}$ ortho-to-para ratios and their uncertainties are plotted for each core in blue. The median otp value as well as its standard deviation for all of the cores (3.5 $\pm$ 0.4) are plotted with a gray horizontal dashed line and gray band, respectively. The statistical value of 3 is marked by a horizontal solid blue line. }
\label{otpf}
\end{figure*}

Three ortho and three para-c-C$_{3}$H$_{2}$ transitions have been detected towards all of the cores (except for Per752 and L1448) which allows us to obtain otp ratios for 14 cores. The ortho-c-C$_{3}$H$_{2}$ column density is divided by the para-c-C$_{3}$H$_{2}$ column density. The otp ratios are listed in Table \ref{otpratt}. In Figure \ref{otpf} the otp ratios and its errors are plotted for the different cores in blue. The median value of 3.5 $\pm$ 0.4 is plotted with a gray dashed line and with a gray band, respectively. The median uncertainty is calculated with the bootstrap method, which is a statistical resampling technique used to estimate the uncertainty of measured parameters. \

All of the cores present a statistical otp ratio of 3 within errors except for core Per799 which has a higher otp ratio: 4.7 $\pm$ 1.6.\

\begin{figure*}[h]
\centering
\includegraphics[width=\textwidth]{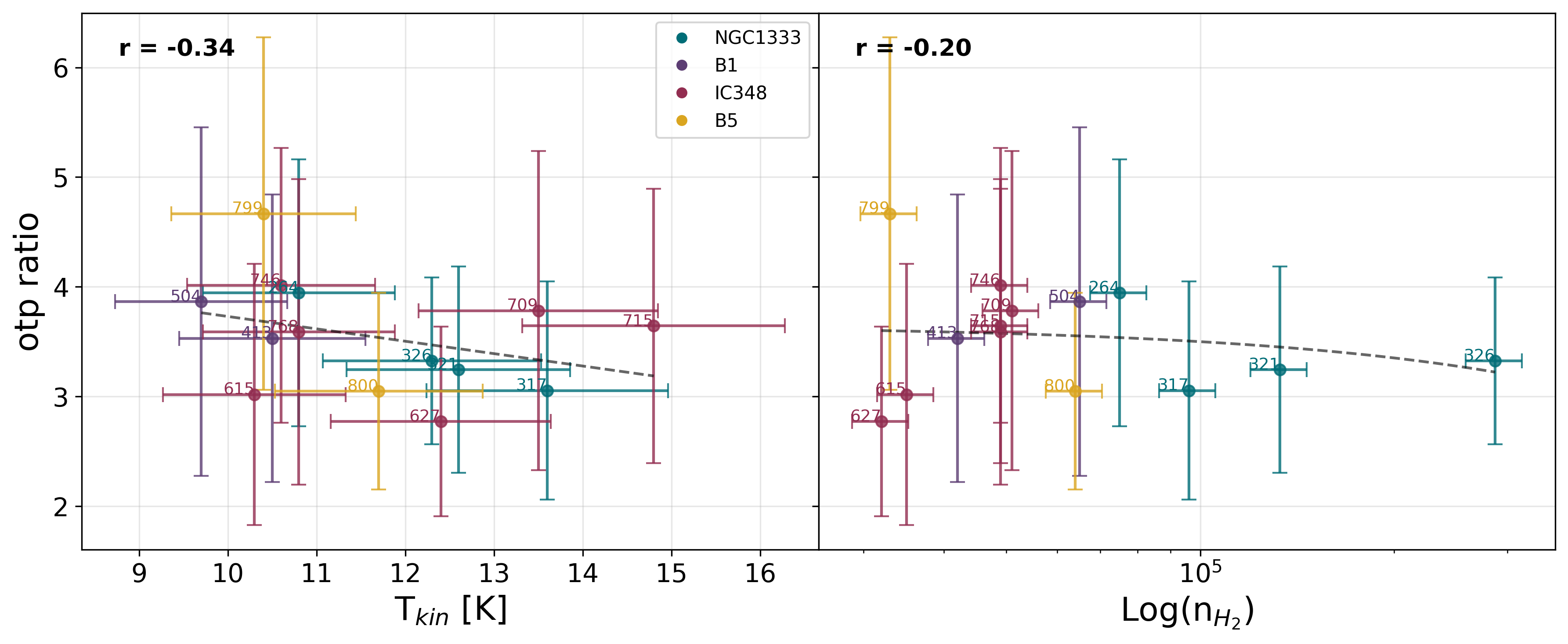}
\caption{Ortho-to-para ratios plotted versus the kinetic temperature, left panel, and volume density, right panel, of the cores. The cores belonging to different regions within the Perseus Molecular Cloud are plotted with different colours (NGC1333 in teal, B1 in purple, IC348 in red and B5 in yellow). The x axis in the right panel is plotted in logarithmic scale for clarity. The gray dashed line indicates the linear correlation trend between the two variables. On the top left corner the Pearson correlation coefficient is displayed. }
\label{otpvsnh2vstkin}
\end{figure*}

The c-C$_{3}$H$_{2}$ otp ratio has been plotted vs the volume density and kinetic temperature of the cores with the aim of finding possible correlations (Figure \ref{otpvsnh2vstkin}, right and left panels respectively). No correlations have been found for the otp ratio either with $n_{\rm H_2}$ nor with $T_{\rm kin}$ ($r$ = -0.20 and -0.34, respectively). Also, no trends have been found for the otp ratio regarding the cloud subregion.\

\subsection{c-C$_{3}$HD} \label{cc3hd}

For c-C$_{3}$HD a total of five lines were targeted, three at lower frequencies with the Yebes 40m telescope: 2$_{11}$ - 2$_{02}$, 1$_{01}$ - 0$_{00}$ and 1$_{11}$ - 0$_{00}$; one at intermediate frequencies with the IRAM 30m telescope: 3$_{03}$ - 2$_{02}$; and two at higher frequencies with the ARO 12 m telescope: 4$_{14}$ - 3$_{03}$ and 2$_{20}$ - 1$_{11}$. The observed lines with their Gaussian fits from pyspeckit are plotted in Figures \ref{occ3hd} and \ref{ol1448} for source L1448. \

c-C$_{3}$HD is detected towards 15/16 cores ($\sim$ 94\%). The only core where c-C$_{3}$HD was not detected is Per752. The column density values can be found on Table \ref{coldenst}, except for Per752 for which we give upper limits. The Per752 core was exclusively observed with the Yebes 40m data, and thus, the upper limit column density in this case is estimated just with the 2$_{1,1}$ - 2$_{0,2}$, 1$_{0,1}$ - 0$_{0,0}$ and 1$_{1,1}$ - 0$_{0,0}$ transitions. A line width of 0.5 km s$^{-1}$ was assumed. For cores Per264, Per317, Per321, Per326, Per709, Per715 and Per800, transition 1$_{1,1}$ - 0$_{0,0}$ is badly fit by RADEX, giving unexpected high $T_{\rm ex}$ values. Thus, this transition is excluded from the fit towards these cores. The c-C$_{3}$HD column density values found range from  2.0 $\pm$ 0.3$\times 10^{11}$ cm$^{2}$(Per317) to 1.9 $\pm$ 0.2$\times 10^{12}$ cm$^{2}$  (Per800), with a weighted average of 4.0 $\pm$ 0.1$\times 10^{11}$ cm$^{2}$ (not taking into account the upper limit of Per752).\

The median of $T_{\rm ex}$ for 2$_{1,1}$ - 2$_{0,2}$ is 8.4 K, for 1$_{0,1}$ - 0$_{0,0}$ is 5.6 K, for 1$_{1,1}$ - 0$_{0,0}$ is 12.8 K, for 4$_{1,4}$ - 3$_{0,3}$ is 4.9 K and for 2$_{2,0}$ - 1$_{1,1}$ is 4.6 K. The optical depth ($\tau$) does not change much within transitions with a median of 0.02.\

\subsection{c-C$_{3}$D$_{2}$} \label{cc3d2}
For c-C$_{3}$D$_{2}$ a total of three lines were targeted, two at lower frequencies with the Yebes 40m telescope: 2$_{11}$ - 2$_{02}$ (o) and 1$_{11}$ - 0$_{00}$ (o); and one at intermediate frequencies with the IRAM 30m telescope: 3$_{03}$ - 2$_{12}$ (p). The observed lines with their Gaussian fits from pyspeckit are plotted in Figures \ref{occ3d2} and \ref{ol1448}, for source L1448.\

c-C$_{3}$D$_{2}$ was detected towards 9/16 cores ($\sim$ 56\%). The cores where c-C$_{3}$D$_{2}$ is not detected are Per264, Per317, Per326, Per615, Per709, Per752 and Per768. There is not a perfect correlation between the cores with the highest c-C$_{3}$HD column densities and c-C$_{3}$D$_{2}$ detections. For example, c-C$_{3}$D$_{2}$ has not been detected towards Per326, one of the cores with the highest c-C$_{3}$HD column densities. \

All of the detected c-C$_{3}$D$_{2}$ transitions are ortho except for one para transition detected towards L1448. As we do not have both ortho- and para-c-C$_{3}$D$_{2}$ for the same cores we assume a statistical otp ratio of 2 (see Section \ref{otprat}) to calculate the total c-C$_{3}$D$_{2}$ column densities. The total-c-C$_{3}$D$_{2}$ column densities can be found on Table \ref{coldenst}, except for Per264, Per317, Per326, Per615, Per709, Per752 and Per768 for which we give upper limits. For these cores c-C$_{3}$D$_{2}$ was exclusively targeted with the Yebes 40m data, and thus, the upper limit column density in this case is estimated with the 2$_{1,1}$ - 2$_{0,2}$ and 1$_{1,1}$ - 0$_{0,0}$ transitions. As these are ortho transitions the total column density upper limit was calculated assuming an ortho-to-para ratio of 2. As for the other molecules, a line width of 0.5 km s$^{-1}$ was assumed. RADEX is not able to find a solution for the para 3$_{0,3}$ - 2$_{1,2}$ transition observed towards L1448. As there is just one c-C$_{3}$D$_{2}$ transition targeted towards L1448, we calculate the column density of the molecule by assuming an excitation temperature of 5 K, as RADEX found a $T_{\rm ex}$ of $\sim$ 6 K for c-C$_{3}$HD towards the same source. The c-C$_{3}$D$_{2}$ column density values found range from 5.8 $\pm$ 1.6$\times 10^{10}$ (Per715) to 1.6 $\pm$ 0.2$\times 10^{11}$ (Per800), with a weighted average of 9.9 $\pm$ 0.7$\times 10^{10}$ (not taking into account the upper limits of non detections).\

The excitation temperature ($T_{\rm ex}$) of c-C$_3$D$_2$ transitions tends to be higher than that of the other isotopologues. The $T_{\rm ex}$ value derived for the 1$_{1,1}$ - 0$_{0,0}$ transition is in average larger than for 2$_{1,1}$ - 2$_{0,2}$, with medians 13.6 and 8.2 K. The median $\tau$ for 1$_{1,1}$ - 0$_{0,0}$  of all cores is 0.004 and for 2$_{1,1}$ - 2$_{0,2}$ is 0.002.\

\subsection{c-C$_{3}$H$_{2}$ Deuteration Ratios} \label{deutrat}

The c-C$_{3}$HD/c-C$_{3}$H$_{2}$ (D/H) ratio has been obtained towards all of the cores except towards Per752, where neither c-C$_{3}$H$_{2}$ nor c-C$_{3}$HD were detected, and L1448, as we do not have data covering c-C$_{3}$H$_{2}$ transitions. The c-C$_{3}$D$_{2}$/c-C$_{3}$HD (D$_{2}$/D) ratio is obtained towards 9 cores including Per321, Per413, Per504, Per627, Per715, Per746, Per799, Per800 and L1448 (see Table \ref{coldenst}). \

Two types of deuteration ratios are given, the direct ones where the column densities of the isotopologues are divided, and the corrected ones, which take into account statistical corrections. We use the statistical correction formula taken from Appendix C of \cite{drozdovskaya:22}. To calculate the (D/H)$_{\rm c-C_3H_2}$ from the c-C$_{3}$HD/c-C$_{3}$H$_{2}$ ratio we employ the following equation, 
\begin{equation}
    \frac{XH_{n-i}D_{i}}{XH_{n}}=\binom{n}{i} \left(\frac{D}{H}\right)^{i}_{XH_{n}}
\end{equation}
where $n$ are the equivalent H atom positions and $i$ is the number of deuteriums. If we substitute and solve for the statistically-corrected ratio (D/H)$_{\rm c-C_3H_2}$ we get,
\begin{equation}
    \left(\frac{D}{H}\right)_{c-C_3H_2} = \frac{1}{2} \times \frac{c-C_{3}HD}{c-C_{3}H_{2}}
\end{equation}
which indicates that the ratio of the c-C$_{3}$HD/c-C$_{3}$H$_{2}$ column densities needs to be multiplied by 1/2. Using the formula,
\begin{equation}
    \frac{XH_{n-i}D_{i}}{XH_{n-j}D_{j}}=\frac{\binom{n}{i}}{\binom{n}{j}} \left(\frac{D}{H}\right)^{i-j}_{XH_{n-j}D_{j}}
\end{equation}
where $j$ is the number of deuteriums in the other isotopologue, we see that to obtain (D/H)$_{c-C_{3}HD}$ we need to multiply c-C$_{3}$D$_{2}$/c-C$_{3}$HD by a factor of 2.\

With the aim of comparing these deuteration ratios with other molecules, we will use the statistically corrected values in discussion.   \

The c-C$_{3}$H$_{2}$ D/H ratios range from 0.5 to 9.2\% with a median value of 1.5 $\pm$ 0.2 \%. The core with the lowest D/H ratio is Per768 and the one with the highest is Per326. The c-C$_{3}$H$_{2}$ D$_{2}$/D ratios range from 9.0 to 55.2\% with a median value of 25.9 $\pm$ 4.3 \%. The core with the lowest detected ratio is Per413 and the one with the highest, Per715. \ 

\begin{figure*}[h]
\centering
\includegraphics[width=\textwidth]{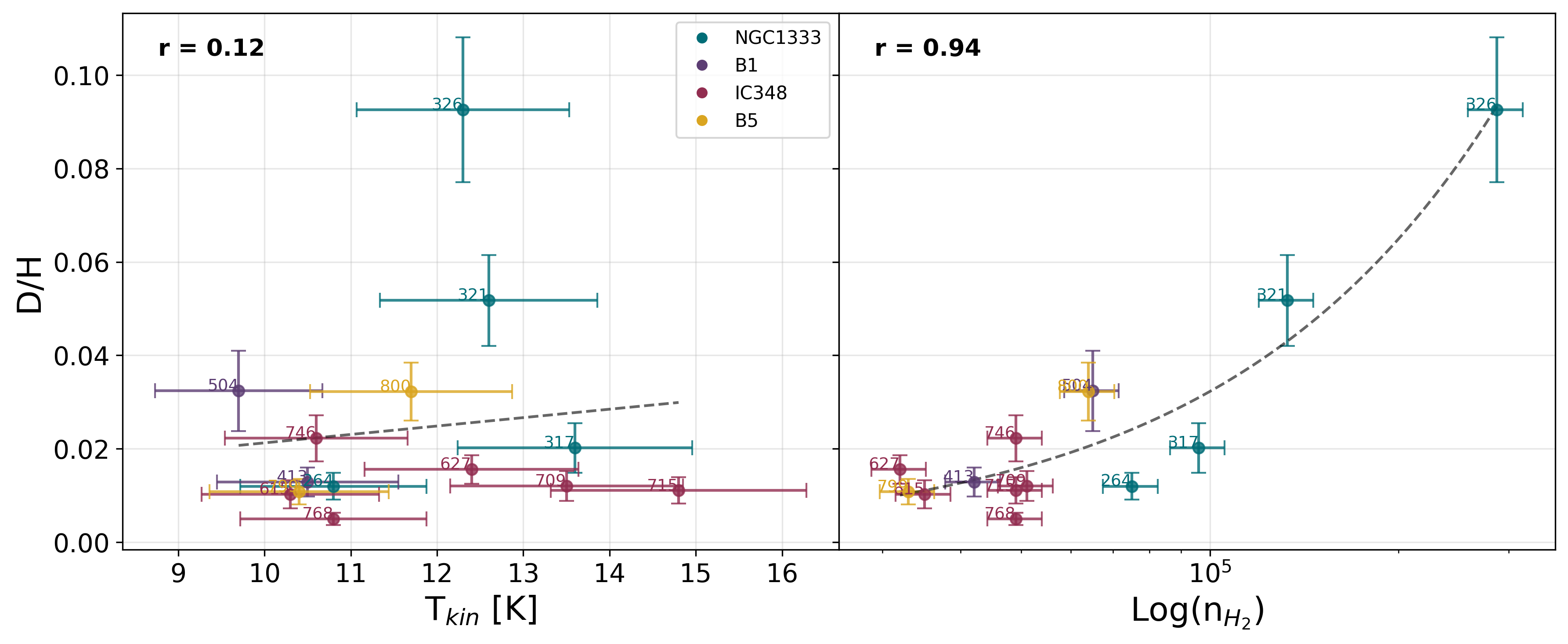}
\caption{D/H ratios plotted versus the kinetic temperature, left panel, and volume density, right panel, of the cores. The cores belonging to different regions within the Perseus Molecular Cloud are plotted with different colours. The x axis in the volume density plot is plotted in logarithmic scale for clarity. The gray dashed line indicates the linear correlation trend between the two variables. The Pearson correlation coefficient is displayed on the top left. }
\label{dvsnh2vstkin}
\end{figure*}

\begin{figure*}[h]
\centering
\includegraphics[width=\textwidth]{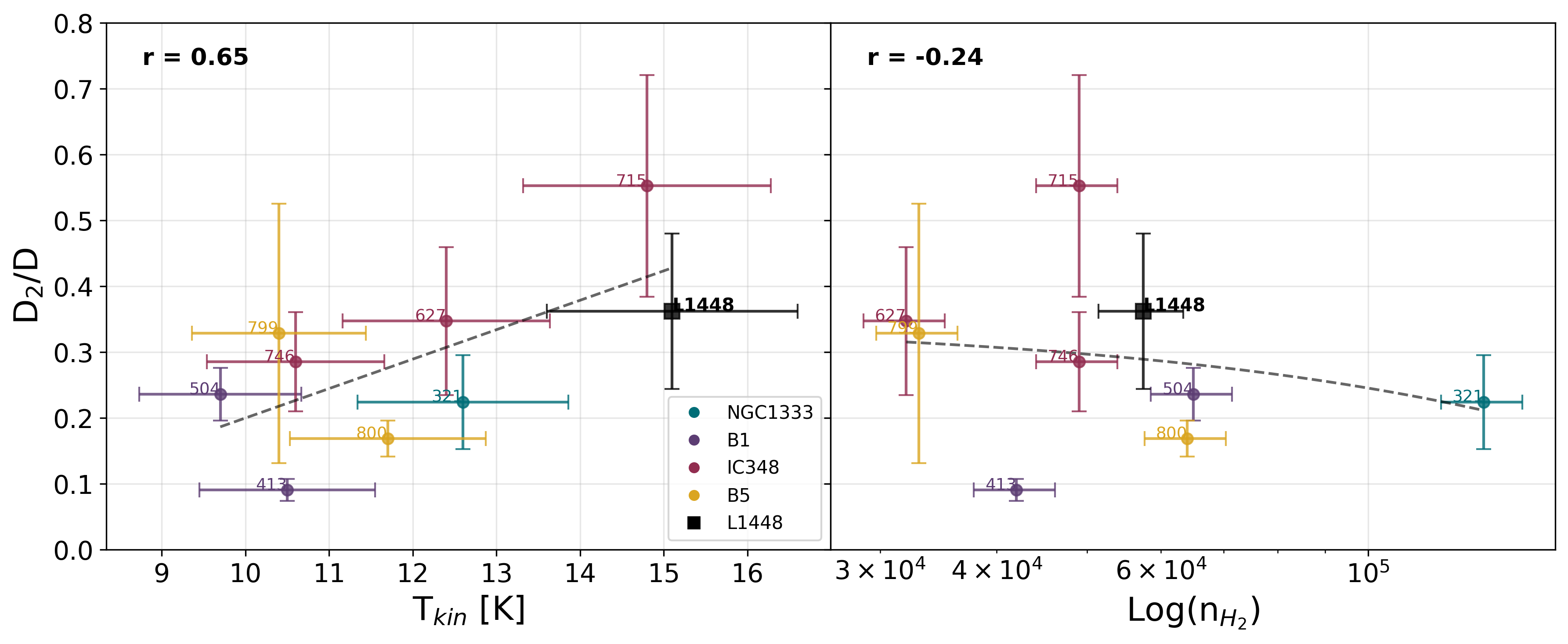}
\caption{D$_{2}$/D ratios plotted versus the kinetic temperature, left panel, and volume density, right panel, of the cores. The cores belonging to different regions within the Perseus Molecular Cloud are plotted with different colours. The x axis in the volume density plot is plotted in logarithmic scale for clarity. The gray dashed line indicates the linear correlation trend between the two variables. The Pearson correlation coefficient is displayed on the top left. }
\label{d2vsnh2vstkin}
\end{figure*}

\begin{figure}[h]
\centering
\includegraphics[width=8cm]{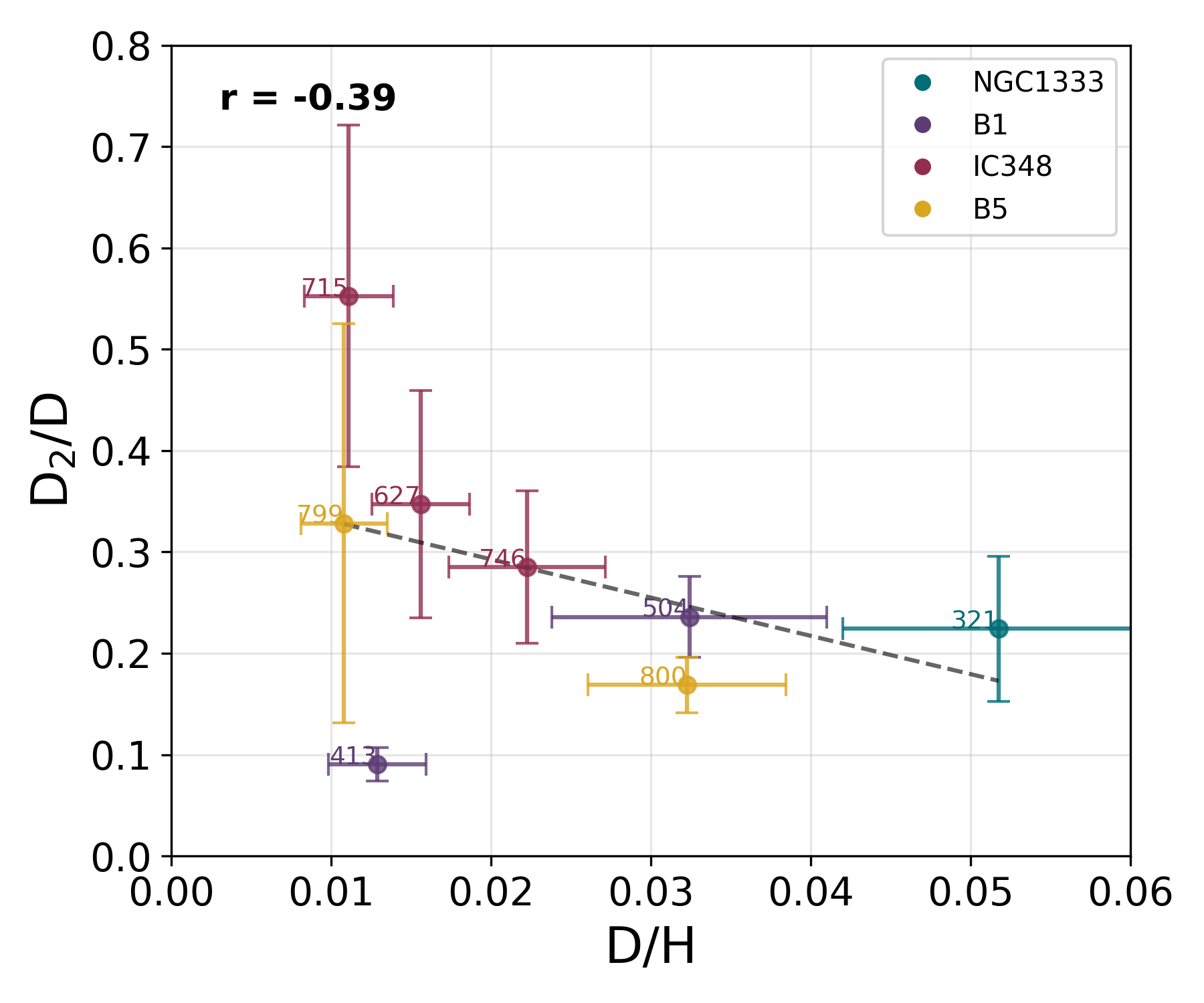}
\caption{D$_{2}$/D ratios plotted versus the D/H ratios. The cores belonging to different regions within the Perseus Molecular Cloud are plotted with different colours.  The gray dashed line indicates the correlation trend between the two variables. On the top left corner the correlation coefficient is displayed.}
\label{dvsd2}
\end{figure}

To explore the possible correlations of deuteration with the physical properties of cores, the D/H and D$_{2}$/D ratios are plotted against the $n_{\rm H_2}$ and $T_{\rm kin}$ (see Figures \ref{dvsnh2vstkin} and \ref{d2vsnh2vstkin}, respectively). A correlation is found between the D/H ratio and the $n_{\rm H_2}$ of the cores, with a correlation coefficient of 0.94, indicating that the D/H ratio is indeed higher for denser cores (Figure \ref{dvsnh2vstkin}, right panel). On the other hand, there is no correlation found for D/H vs $T_{\rm kin}$ ($r$ = 0.12, Figure \ref{dvsnh2vstkin}, left panel). Contrary to the D/H ratio, D$_{2}$/D presents no correlation with $n_{\rm H_2}$ ($r$ = -0.24, Figure \ref{d2vsnh2vstkin}, right panel), but it shows a correlation with $T_{\rm kin}$ ($r$ = 0.65, \ref{d2vsnh2vstkin}, left panel). Finally, plotting D/H against D$_{2}$/D does not give a strong correlation ($r$ = -0.39, Figure \ref{dvsd2}).\

\begin{figure}[h]
\centering
\includegraphics[width=8cm]{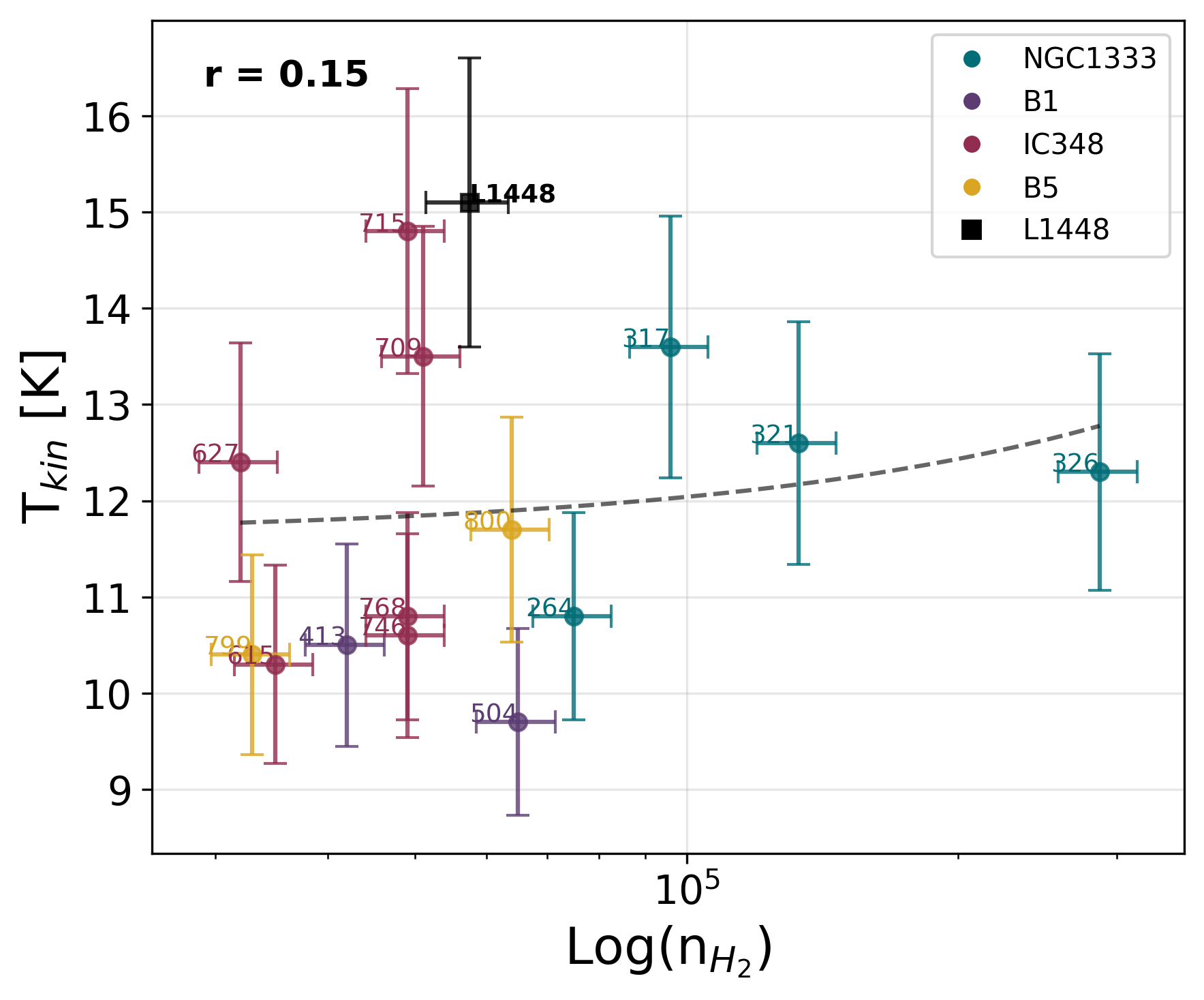}
\caption{Kinetic temperatures of clouds plotted against their volume densities. The x axis is plotted in logarithmic scale for clarity. The cores belonging to different regions within the Perseus Molecular Cloud are plotted with different colours. The gray dashed line indicates the correlation trend between the two variables. On the top right corner the correlation coefficient is displayed.}
\label{nh2vstkin}
\end{figure}

\begin{table*}[ht]
\begin{center}
\caption{Obtained column densities, D/H and D$_{2}$/D ratios. }

\resizebox{\textwidth}{!}{%
\begin{tabular}{ lccccccc } 
\hline\hline
Core Number & c-C$_{3}$H$_{2}$ & c-C$_{3}$HD & c-C$_{3}$D$_{2}$ & D/H & D/H$_{\rm corr}$ & D$_{2}$/D & D$_{2}$/D$_{\rm corr}$   \\
\hline
264 &  1.8 $\pm$ 0.4$\times 10^{13}$ & 4.4 $\pm$ 0.6$\times 10^{11}$  & $\leq$5.7 $\pm$ 1.4$\times 10^{10}$ & 0.024(0.006) & 0.012(0.003) & - & - \\
317 &  5.0 $\pm$ 1.0$\times 10^{12}$ & 2.0 $\pm$ 0.3$\times 10^{11}$  & $\leq$6.6 $\pm$ 1.6$\times 10^{10}$ & 0.040(0.010)& 0.020(0.005) & -  & - \\
321 &  6.3 $\pm$ 1.1$\times 10^{12}$ & 6.6 $\pm$ 0.5$\times 10^{11}$  & 7.4 $\pm$ 2.3$\times 10^{10}$ & 0.103(0.019) & 0.052(0.010) & 0.112(0.036)    & 0.224(0.071) \\
326 &  8.0 $\pm$ 1.1$\times 10^{12}$ & 1.5 $\pm$ 0.1$\times 10^{12}$  & $\leq$5.4 $\pm$ 1.4$\times 10^{10}$ & 0.185(0.031) & 0.092(0.015) & - & - \\
413 &  8.1 $\pm$ 1.8$\times 10^{13}$ & 2.1 $\pm$ 0.1$\times 10^{12}$  & 9.5 $\pm$ 1.6$\times 10^{10}$ & 0.026(0.006) & 0.013(0.003) & 0.045(0.008)    & 0.090(0.017) \\
504 &  1.8 $\pm$ 0.4$\times 10^{13}$ & 1.1 $\pm$ 0.1$\times 10^{12}$ & 1.3 $\pm$ 0.2$\times 10^{11}$ & 0.065(0.017) & 0.032(0.009) & 0.118(0.020)    & 0.236(0.040) \\
615 &  2.1 $\pm$ 0.5$\times 10^{13}$ & 4.3 $\pm$ 0.6$\times 10^{11}$  & $\leq$5.1 $\pm$ 1.2$\times 10^{10}$ & 0.020(0.006) & 0.010(0.003) & - & - \\
627 &  1.7 $\pm$ 0.3$\times 10^{13}$ & 5.4 $\pm$ 0.5$\times 10^{11}$  & 9.3 $\pm$ 2.9$\times 10^{10}$ & 0.031(0.006) & 0.016(0.003)& 0.173(0.056)  & 0.347(0.112)  \\
709 &  1.6 $\pm$ 0.4$\times 10^{13}$ & 4.0 $\pm$ 0.6$\times 10^{11}$  & $\leq$4.6 $\pm$ 1.7$\times 10^{10}$ & 0.024(0.006) & 0.012(0.003) & - & - \\
715 &  9.4 $\pm$ 2.1$\times 10^{12}$ & 2.1 $\pm$ 0.2$\times 10^{11}$  & 5.8 $\pm$ 1.6$\times 10^{10}$ & 0.022(0.006) & 0.011(0.003) & 0.276(0.084)  & 0.552(0.168) \\
746 &  1.6 $\pm$ 0.3$\times 10^{13}$ & 7.2 $\pm$ 0.9$\times 10^{11}$ & 1.0 $\pm$ 0.2$\times 10^{11}$ & 0.044(0.010) & 0.022(0.005) & 0.142(0.038)    & 0.285(0.075) \\
752 & $\leq$4.2 $\pm$ 1.7$\times 10^{12}$ & $\leq$3.1 $\pm$ 0.8$\times 10^{11}$ & $\leq$5.6 $\pm$ 1.4$\times 10^{10}$ & - & - & - & -   \\
768 &  4.3 $\pm$ 1.0$\times 10^{13}$ & 4.2 $\pm$ 0.5$\times 10^{11}$ & $\leq$5.0 $\pm$ 1.3$\times 10^{10}$ & 0.010(0.003) & 0.005(0.001) & - & - \\
799 &  2.8 $\pm$ 0.6$\times 10^{13}$ & 6.0 $\pm$ 0.6$\times 10^{11}$ & 9.8 $\pm$ 5.8$\times 10^{10}$ & 0.022(0.005) & 0.011(0.003) & 0.164(0.098)    & 0.328(0.197) \\
800 &  3.0 $\pm$ 0.5$\times 10^{13}$ & 1.9 $\pm$ 0.2$\times 10^{12}$ & 1.6 $\pm$ 0.2$\times 10^{11}$ & 0.064(0.012) & 0.032(0.006) & 0.084(0.013)    & 0.169(0.027) \\
L1448 & -  & 7.8 $\pm$ 1.9$\times 10^{11}$ & 1.4 $\pm$ 0.3$\times 10^{11}$ & - & - & 0.181(0.059)    & 0.362(0.118) \\
\hline
\end{tabular}%
}
\tablefoot{The "-" symbol is used to indicate towards the L1448 the c-C$_{3}$H$_{2}$ molecule was not targeted and also that some D/H and D$_{2}$/D have not been able to be calculated due to one or more isotopologues not being detected. The D/H ratios have been corrected by multiplying by two and the D$_{2}$/D by dividing by two. The upper limits correspond to the 3$\sigma$ level.}
\label{coldenst}
\end{center}
\end{table*}

\begin{table}[ht]
\begin{center}
\caption{Obtained c-C$_{3}$H$_{2}$ ortho-to-para ratios along the studied cores.}

\begin{tabular}{ lc } 
\hline\hline
Core Number & otp ratio  \\
\hline
264 &  3.9$\pm$1.2  \\
317 &  3.0$\pm$1.0  \\
321 &  3.2$\pm$0.9  \\
326 &  3.3$\pm$0.8  \\
413 &  3.5$\pm$1.3  \\
504 &  3.9$\pm$1.6  \\
615 &  3.0$\pm$1.2  \\
627 &  2.8$\pm$0.9  \\
709 &  3.8$\pm$1.4  \\
715 &  3.6$\pm$1.2  \\
746 &  4.0$\pm$1.2  \\  
768 &  3.6$\pm$1.4  \\
799 &  4.7$\pm$1.6  \\
800 &  3.0$\pm$0.9  \\
\hline
\label{otpratt}
\end{tabular}
\end{center}
\end{table}

\section{Discussion} \label{discussion}

c-C$_{3}$H$_{2}$ has been detected towards the entire subset of cores, except Per752, for which an upper column density limit is given. Per752 is the core with the lowest volume density of the cores studied in this paper. This, in combination with the slightly larger $T_{\rm kin}$, could result in not sufficient emission to be detected with the sensitivity of the current observations. However, other molecules, including COMs, e.g. CH$_{3}$OH, CH$_{3}$CHO, CH$_{3}$CN, t-HCOOH and H$_{2}$CCO have been detected towards this core \citep{scibelli:24}. Thus, the c-C$_{3}$H$_{2}$, non detection towards Per752 appears to be directly related to this specific molecule instead of this core being generally chemically poor. In Pokorny-Yadav et. al. (in prep.), carbon chains were specifically surveyed in the same starless and prestellar core selection towards the Perseus Molecular Cloud. In the above mentioned study, Per752 presents almost no carbon chains detection, which is in agreement with the non detection of c-C$_{3}$H$_{2}$ in this study. Thus, Per752 is particularly poor in carbon chain molecules, suggesting that the physical characteristics and environment of this core are favourable for the production of COMs but not of carbon chains (see Pokorny-Yadav et. al. (in prep.)).  \ 

\subsection{c-C$_{3}$H$_{2}$ otp Ratio}
The statistical ortho to para ratio of c-C$_{3}$H$_{2}$ is 3 (see Section \ref{otprat}). However, this ratio can deviate from the statistical value due to different factors such as the initial non-statistical otp ratio of the reactants and interconversion ratio reactions \citep{furuya:15, lupi:21}.\  

Observational studies have explored the otp ratio of c-C$_{3}$H$_{2}$ towards TMC-1 \citep{madden:89, takakua:01, morisawa:06} and L1527 \citep{takakua:01}. These studies showed that the otp ratio towards prestellar cores can deviate from the statistical value.  \cite{morisawa:06} find a direct correlation between the core's evolutionary stage, as measured with the NH$_{3}$/CCS abundance ratio, and its otp ratio, with lower otp ratios observed towards younger cores. However, the gas-phase chemical modelling of the otp ratio of c-C$_{3}$H$_{2}$ done in \citealp{park:05}, which includes ortho-to-para interconversion reactions, is not able to reproduce the observed ratios in \cite{morisawa:06}. Nevertheless, they claim observational uncertainties or the incompleteness of the reaction network could have impacted the results. Thus, the correlation between the otp ratio and the evolutionary stage of the core is not confirmed. \

The otp ratio of c-C$_{3}$H$_{2}$ studies are limited to TMC-1 and L1527. In this study, we expand these observations to the Perseus Molecular Cloud to understand whether the deviation of the otp ratio also happens in other starless and prestellar cores.\

As seen in Section \ref{otprat}, all of the cores except for one have a statistical otp ratio of 3 within uncertainty. The median otp ratio of the core sample is 3.5 $\pm$ 0.4. Due to the limited sensitivity of the observations, the uncertainties or the otp ratios are large, and possible non-statistical ratios towards other cores are not seen. Due to the possibility of c-C$_{3}$H$_{2}$ otp ratios deviating from its statistical value of 3, as seen in previous works \citep{madden:89, takakua:01, morisawa:06} and as hinted in this one, calculating total c-C$_{3}$H$_{2}$ column densities from either only the ortho- or para-c-C$_{3}$H$_{2}$ column densities by assuming an otp ratio of 3 is not always safe. If possible, both ortho- and para-c-C$_{3}$H$_{2}$ transitions should be observed to get an accurate total c-C$_{3}$H$_{2}$ column density. Plotting otp ratios against $n_{\rm H_2}$ and $T_{\rm kin}$ does not give any trends (with correlation coefficients of -0.20 and -0.34 in Figure \ref{otpvsnh2vstkin}, right and left panels, respectively). If we assume a direct relationship between the core volume density and its evolutionary stage, contrarily to \cite{morisawa:06}, we do not see a trend of increasing otp ratios with evolutionary stage (Figure \ref{otpvsnh2vstkin}). The absence of a trend could be due to the uncertainties related to the derived otp ratios. However, the c-C$_{3}$H$_{2}$ otp ratios may also be determined by many more variables other than core evolutionary stage, making the c-C$_{3}$H$_{2}$ otp ratio not a reliable proxy for core evolutionary stage.\

\subsection{c-C$_{3}$H$_{2}$ Deuteration}

c-C$_{3}$HD, as its main isotopologue, is detected towards all of the cores in the sample except Per752. On the other hand, c-C$_{3}$D$_{2}$ is detected towards a smaller number of cores (Per321, Per413, Per504, Per627, Per715, Per746, Per799 and Per800). To test whether the cores where c-C$_{3}$D$_{2}$ is detected could be the cores with highest D/H ratios, we plotted the D/H vs the D$_{2}$/D ratios, but no clear trend was seen ($r$ = -0.39, Figure \ref{dvsd2}).\

Plotting the c-C$_{3}$H$_{2}$ D/H and D$_{2}$/D ratios against the volume density ($n_{\rm H_2}$) of the cores shows a direct correlation between D/H and $n_{\rm H_2}$ ($r$ = 0.94, Figure \ref{dvsnh2vstkin}, right panel) but no correlation for D$_{2}$/D and $n_{\rm H_2}$ ($r$ = -0.24, Figure \ref{d2vsnh2vstkin}, right panel). D/H scaling with $n_{\rm H_2}$ agrees with the idea that the more evolved cores, which are denser, are the ones that present higher deuteration ratios. Nevertheless, in \cite{chantzos:18} they do not find a correlation between the D/H ratio and the central column densities of Taurus starless and prestellar cores. If Per326 is removed from the Perseus correlation then a Pearson correlation coefficient of 0.76 is obtained. If both the Per326 and Per321 cores are removed then a Pearson correlation coefficient of 0.42 is obtained. If instead the Spearman rank correlation coefficient ($\rho$), which measures the strength and direction of a monotonic relationship between two variables, is used, the value 0.64 is obtained. The correlation found between D/H and $n_{\rm H_2}$ seem to be governed by these two more massive cores. On the other hand, the D$_{2}$/D ratio does not appear to follow this trend. This can be due to the fact that the detection sample spans a narrower range in volume densities compared to the D/H ratio. Nevertheless, the D$_{2}$/D ratio not correlating with volume density may point to the independence of the second deuteration of c-C$_{3}$H$_{2}$ from density, unlike D/H ratio.\

The c-C$_{3}$H$_{2}$ D/H and D$_{2}$/D ratios were also plotted against the kinetic temperature ($T_{\rm kin}$) of the cores, as it is known that low temperatures enhance deuterium fractionation (Figures \ref{dvsnh2vstkin} and \ref{d2vsnh2vstkin}, respectively). No clear trend is observed for the D/H ratio, but there is a positive correlation for the D$_{2}$/D one. The lack of correlation between D/H and $T_{\rm kin}$ may be due to the fact that $T_{\rm kin}$ does not necessarily represent the temperature of the inner part of the core where deuterium fractionation is enhanced. This is supported by the lack of correlation ($r$ = 0.15) between $n_{\rm H_2}$ and $T_{\rm kin}$ (Figure \ref{nh2vstkin}), which would be expected if $T_{\rm kin}$ represented the temperature at the centre of the core which would decrease with increasing $n_{\rm H_2}$. The positive correlation of D$_{2}$/D with $T_{\rm kin}$ could be explained if the second deuteration happened, at least partly, in the gas phase through slightly endothermic reactions, as for example with CH$_{2}$D$^{+}$. \

c-C$_{3}$H$_{2}$ is thought to be formed in the gas phase mainly by the electron recombination of c-C$_{3}$H$_{3}^{+}$ \citep{loison:17}. The formation of its deuterated isotopologues is thought to occur through the deuteration of its main isotopologue with H$_{2}$D$^{+}$, D$_{2}$H$^{+}$ and D$_{3}^{+}$, followed by an electron recombination \citep{spezzano:13}. In this scenario, the deuteration fraction increases with time, owing to the cores getting lower temperatures at their centre due to their increasing densities, and the H$_{2}$D$^{+}$ molecule being formed more efficiently. This is seen in our results where the D/H ratio directly correlates with the core volume density. On the other hand we see that the D$_{2}$/D ratio does not correlate with $n_{\rm H_2}$, but positively correlates with $T_{\rm kin}$. This seems to indicate the second deuteration of c-C$_{3}$H$_{2}$ is enhanced by a slightly endothermic reaction. Then when $T_{\rm kin}$ is higher it favours these type of reactions resulting into higher D$_{2}$/D ratios. Reaction with CH$_{2}$D$^{+}$ is thought to become the main deuteration path at warm temperatures ($\gtrsim$30 K), as the formation mechanism of CH$_{2}$D$^{+}$, is slightly more endothermic than the formation mechanism of H$_{2}$D$^{+}$. It is possible then, that at the "warmer" external layers of the cores, traced by $T_{\rm kin}$, deuteration reactions with CH$_{2}$D$^{+}$ start being relevant. There is also the possibility that other slightly endothermic reactions, are responsible for the higher D$_{2}$/D ratios observed towards warmer cores in the Perseus Molecular Cloud. Follow-up work would be needed to assess this scenario in greater detail. \  

The larger D$_{2}$/D ratio compared to the D/H one indicates that the second deuteration, from c-C$_{3}$HD to c-C$_{3}$D$_{2}$, is more effective than the first deuteration from c-C$_{3}$H$_{2}$ to c-C$_{3}$HD. This phenomenon, where the second deuteration is more effective than the first one, has been observed in other molecules. For example, observations towards the prestellar core L1544 show that D$_2$CS is relatively more abundant than HDCS, indicating enhanced second deuteration of thioformaldehyde \citep{spezzano:22}. Studies of formaldehyde towards the protostellar system IRAS~16293-2422 reveal a higher D$_2$CO/HDCO ratio than HDCO/H$_2$CO \citep{jorgensen:18}; similarly, the amidogen radical in the same source exhibits unexpectedly high ND$_2$/NHD ratios, suggesting efficient multiple deuteration during prolonged cold prestellar phases \citep{melosso:20}. This trend has also been observed for some COMs in IRAS~16293-2422 and the neighbouring prestellar core IRAS~16293E (\cite{scibelli:25}, for D$_{2}$/D in methanol). COMs in IRAS~16293-2422 display an average D$_{2}$/D ratio of $\sim$ 20$\%$ which is generally four times higher than their D/H ratios \citep{manigand:20, richard:21, drozdovskaya:22, ferrerasensio:22}. These results point at least to the partial inheritance of these molecules from the prestellar phase where the conditions favour deuterium fractionation. This is confirmed by the methanol D/H and D$_{2}$/D ratios derived from the prestellar core IRAS~16293E \citep{scibelli:25} which formed in the same environment as the protostellar system IRAS~16293-2422.\  

It needs to be taken into account that some c-C$_{3}$H$_{2}$ transitions are optically thick $\tau$ $>$ 1 towards some cores Per413, Per504, Per627, Per768, Per799 and Per800. This means some D/H ratios could be enhanced as the column density of c-C$_{3}$H$_{2}$ would be underestimated. Nevertheless, these cores are not the densest cores that seem to be dominating the trend.\

In Pokorny-Yadav et. al. (in prep.) they have studied the deuteration of HC$_{3}$N across the same group of Perseus starless and prestellar cores, finding a D/H ratio ranging between 1.8\% (Per326) and 12.4\% (Per321), with a median of 6.4\%. In the case of c-C$_{3}$H$_{2}$ the core with the lowest D/H ratio is Per768 (0.5\%), and the one with the highest is Per326 (9.2\%), with a median of 1.5\%. From these results we see that the formation of the singly-deuterated species for HC$_{3}$N is more efficient than the one for c-C$_{3}$H$_{2}$. Pokorny-Yadav et. al. (in prep.) have found a correlation with the volume density of the cores, similarly as found for the D/H ratio of c-C$_{3}$H$_{2}$ in the present study (Figure 12 in Pokorny-Yadav et. al., in prep.). Nevertheless, in the case of HC$_{3}$N, Per326 and Per264 do not follow the trend, having lower D/H ratios. This behaviour is attributed to the local environment of cores Per326 and Per264, affecting the deuteration of HC$_{3}$N. These cores are located on the path of the outflows of the SVS13, IRAS 4A/B and IRAS 2A/B protostars. They suggest that shocks produced by the outflow impacts may release CO from dust grains, lowering the abundance of gas-phase deuteration molecules such as H$_{2}$D$^{+}$. Contrarily to what is found for HC$_{3}$N, the c-C$_{3}$H$_{2}$ D/H ratio of cores Per264 and Per326 follow the positive correlation with $n_{\rm H_2}$. This difference may come from the different chemical nature of the two molecules and their reactivity. Lastly, the HC$_{3}$N D/H median found towards the Perseus starless and prestellar cores in Pokorny-Yadav et. al. (in prep.) is consistent with D/H values measured in other prestellar and protostellar cores in Taurus and Serpens.\

Moreover, the deuteration of H$_{2}$S, H$_{2}$CS and CH$_{3}$OH have also been studied towards some starless and prestellar cores in the Perseus Molecular Cloud. The deuteration of H$_{2}$S has been studied towards ten Perseus starless and prestellar cores, alongside cores in Taurus and Orion, in \cite{rodriguez:23}. HDS has been detected towards seven cores, while D$_{2}$S has been detected towards four cores. The D/H and D$_{2}$/D ratios derived towards the sample is higher than the ratios derived for c-C$_{3}$H$_{2}$ in this study. The deuteration of H$_{2}$CS is studied towards ten starless and prestellar cores in Perseus, additionally to other sources in Taurus and Orion. HDCS is detected towards seven cores, while D$_{2}$CS is detected towards five sources. The derived deuteration H$_{2}$CS ratios, appear higher than the ones found for c-C$_{3}$H$_{2}$ in this study. The higher deuteration ratios of H$_{2}$CS with respect to c-C$_{3}$H$_{2}$ was also observed towards the prestellar core L1544 \citep{spezzano:22}. The D/H ratio of CH$_{3}$OH has been studied towards L1448 in \cite{kulterer:25} giving a higher value also than the median D/H found in Perseus for c-C$_{3}$H$_{2}$. In general, previous studies have shown that H$_{2}$CS and H$_{2}$CO exhibit higher deuteration ratios towards starless and prestellar cores than c-C$_{3}$H$_{2}$, H$_{2}$S, H$_{2}$O and CH$_{3}$OH, which highlights the diversity in deuteration fractionation mechanisms for different molecules.\

\subsubsection{Comparison with the Taurus and Chamaeleon Molecular Clouds} \label{comparison}

The c-C$_{3}$H$_{2}$ D/H and D$_{2}$/D ratios have been derived towards other starless and prestellar cores in previous works. Most of these target cores in the Taurus Molecular Cloud except one work that derives the D/H value towards the prestellar core Cha-C2 in the Chamaeleon Molecular Cloud \citep{lis:25}. Thus, a comparison of the deuteration of c-C$_{3}$H$_{2}$ in starless and prestellar cores between the Perseus, Taurus and Chamaeleon Molecular Clouds can be made. The data for Taurus was taken from \cite{spezzano:13, chantzos:18} and \cite{giers:22}. As the prestellar core L1544 appears in the three papers, we use the c-C$_{3}$H$_{2}$ D/H and D$_{2}$/D ratios derived from the latest work \citep{giers:22}. The ratios in these papers are not statistically corrected. Thus, for comparison purposes, we applied the same statistical factors used in this study (see Section \ref{deutrat}).\

\begin{figure*}[ht]
\centering
\includegraphics[width=19cm]{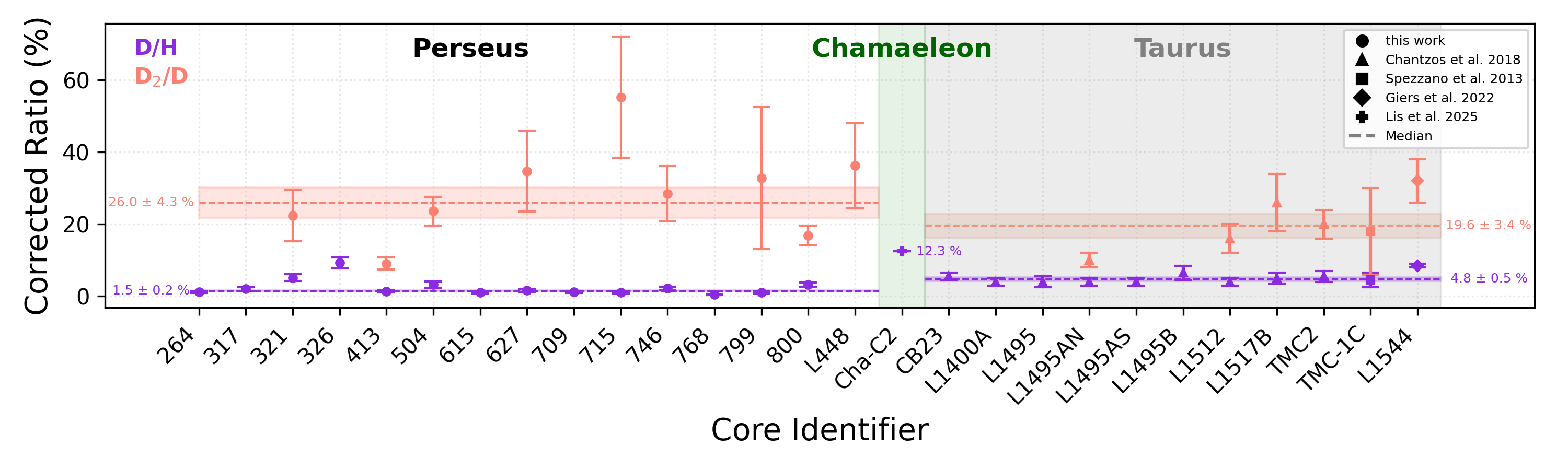}
\caption{Statistically corrected D/H and D$_{2}$/D ratios percentages for c-C$_{3}$H$_{2}$, in purple and orange, respectively, for the sources in the Perseus Molecular Cloud (this study, circles); and additional sources in the Chamaeleon Molecular Cloud: Cha-C2 (\cite{lis:25}, cross); Taurus Molecular Cloud, shaded in gray: CB23, L1400A, L1495, L1495AN, L1495AS, L1495B, L1512, L1517B and TMC2 (\cite{chantzos:18}, triangles); TMC-1C (\cite{spezzano:13}, square) and L1544 (\cite{giers:22}, diamond). The horizontal purple and orange dashed lines and bands, indicate the median with standard deviation for the D/H and D$_{2}$/D for the Perseus and the Taurus cores separately. For the Chamaeleon Molecular Cloud just one source, Cha-C2, is reported, so no median is given. Instead the value of the corrected D/H ratio is plotted.  }
\label{deutf}
\end{figure*}

All the ratios are plotted in statistically corrected percentages (Figure \ref{deutf}). For each singly- and doubly-deuterated ratio and molecular cloud a median value with uncertainties is computed. The uncertainties are computed with the same method used to calculate the otp ratio median uncertainty, and are plotted with horizontal dashed lines and bands. The Perseus median D/H value is 1.5 $\pm$ 0.2 \%, and the D$_{2}$/D median value is 25.9 $\pm$ 4.3 \%. The Taurus median D/H value is 4.8 $\pm$ 0.5 \%, and the D$_{2}$/D median value is 19.6 $\pm$ 3.4 \%. As for Chamaeleon there is just one measurement instead of the median, the corrected D/H value (12\%) is given. The cores in Taurus have a larger c-C$_{3}$H$_{2}$ D/H ratio compared to the Perseus ones. The Cha-C2 prestellar core has a D/H value higher (12\%) than the sources both in Perseus and Taurus. If the correlation between D/H and $n_{\rm H_2}$ is also present in the Taurus and Chamaeleon Molecular Clouds, the higher D/H ratio of Taurus and Chamaeleon (Cha-C2) compared to the one in Perseus would make sense as the cores in Taurus and in Chamaeleon have a higher median $n_{\rm H_2}$ (1.4 $\times$ 10$^{5}$ \footnote{The Taurus starless and prestellar core volume densities are calculated in the same way as for the Perseus sources.} and 5.5 $\times$ 10$^{5}$ cm$^{-3}$, \cite{chantzos:18, belloche:11}) compared to the cores in Perseus. The D$_{2}$/D ratios are equivalent between the Perseus and Taurus Molecular Clouds. These results show there is greater variation of D/H and D$_{2}$/D ratios among cores within the same molecular cloud, than among the medians of different molecular clouds. Which suggests the local environment of the cores has a greater effect on the deuteration of c-C$_{3}$H$_{2}$ than the parent molecular cloud they are located in. \ 

\subsubsection{Comparison with Protostars}\label{protostars}

Besides studying the c-C$_{3}$H$_{2}$ deuteration towards starless and pre-stellar cores towards the Taurus Molecular Cloud in \cite{chantzos:18}, they also studied c-C$_{3}$HD and c-C$_{3}$D$_{2}$ towards some protostellar sources in the Taurus (L1521F) and Perseus (Per5, IRAS03282, HH211 and L1448-IRS2) Molecular Clouds. The mean derived D/H ratio towards the protostellar sources in the Perseus Molecular Cloud is 7 $\pm$ 1 \%, derived taking into account all Perseus sources mentioned above, and the D$_{2}$/D ratio is 34 $\pm$ 6 \%, derived from Per5 and HH211. These ratios take into account statistical corrections. The protostellar D/H median is higher than the starless and prestellar one derived in this study. This is likely due to the prestellar core that evolved into the observed protostars being longer lived than the starless and prestellar cores studied in this study. The protostellar D$_{2}$/D median is also higher than the one for starless and prestellar cores, but they agree within errors. They derived D/H and D$_{2}$/D ratios towards the L1521F protostellar core (Very Low Luminosity Object, or VeLLO) in the Taurus Molecular Cloud of 4 $\pm$ 2 \% and 16 $\pm$ 4 \%. These ratios agree within errors with the mean values derived for starless and prestellar cores in Taurus. The protostellar D/H and D$_{2}$/D ratios generally agree or are higher than the observed starless and prestellar ratios, suggesting at least a partial inheritance of the c-C$_{3}$H$_{2}$ budget, and supporting the idea that the deuterium fractionation increases throughout the prestellar core phase. \ 

\section{Conclusions}\label{conclusions}
This work has presented new c-C$_{3}$H$_{2}$, c-C$_{3}$HD and c-C$_{3}$D$_{2}$ observations towards starless and prestellar cores towards the Perseus Molecular Cloud taken with the Yebes 40m, the ARO 12m and the IRAM 30m telescopes. c-C$_{3}$H$_{2}$ and c-C$_{3}$HD are detected towards all targeted cores except for Per752. c-C$_{3}$D$_{2}$ is detected towards 56\% (9/16) of the targeted cores.\

The ortho-to-para ratio of c-C$_{3}$H$_{2}$ has been studied leading to a median value of 3.5 $\pm$ 0.4 for all targeted cores. All targeted cores but Per799 (with a value of 4.7 $\pm$ 1.6) have a statistical otp ratio within uncertainties. No correlation has been found between c-C$_{3}$H$_{2}$ otp ratios and the evolutionary stage of the core, traced by $n_{\rm H_2}$.\

The median statistically corrected c-C$_{3}$H$_{2}$ D/H and D$_{2}$/D ratios derived from the observations are 1.5 $\pm$ 0.2 \% and 25.9 $\pm$ 4.3 \%. No correlation has been found between D/H and D$_{2}$/D ratios. On the other hand, D/H ratios positively correlate with the core's volume density. This supports the idea that more evolved, and therefore, more dense cores, present higher levels of deuteration. The D$_{2}$/D ratios positively correlate with $T_{\rm kin}$, suggesting that, while the first deuteration is dependent on volume density, the second deuteration is dependent on $T_{\rm kin}$ and may happen through slightly endothermic reactions. The median D/H ratio in Perseus is lower than the observed in cores towards the Taurus Molecular Cloud and the core observed towards the Chamaeleon Molecular Cloud. This could be attributed to the cores in the latter Molecular Clouds having higher $n_{\rm H_2}$ than the ones in Perseus. When it comes to D$_{2}$/D, even if the median found in Perseus is larger than in Taurus, they are equivalent taking uncertainties into account. \

The variation of D/H and D$_{2}$/D ratios among cores is larger than the variation of their medians among molecular clouds, suggesting that the local environment of the cores have a higher impact on the deuteration of c-C$_{3}$H$_{2}$ than the parent cloud itself.\

\textit{Acknowledgements.} J. Ferrer Asensio thanks RIKEN Special Postdoctoral Researcher Program (Fellowships) for financial support. S. Scibelli acknowledges the National Radio Astronomy Observatory which is a facility of the National Science Foundation operated under cooperative agreement by Associated Universities, Inc. L. Steffes is supported by a National Science Foundation Graduate Research Fellowship Program under Grant \textnumero [DGE-2137419]. Any opinions, findings, and conclusions or recommendations expressed in this material are those of the authors and do not necessarily reflect the views of the National Science Foundation. L. Steffes and Y. Shirley are also supported by the National Science Foundation Astronomy and Astrophysics Grant (AAG) AST-2205474. We are thankful to have the oportunity to conduct astronomical research on Iolkam Du'ag (Kitt Peak) in Arizona and we recognize and acknowledge the significant cultural role and reverence that these sites have to the Tohono O'odham Nation. We sincerely thank the operators of the Arizona Radio Observatory (Kevin Bays, Patrick Fimbres, Blythe Guvenen, and Ryan Buchta) for their assistance with observations. The 12m Telescope is operated by the Arizona Radio Observatory (ARO), Steward Observatory, and the University of Arizona, with funding from the State of Arizona, NSF MRI Grant AST-1531366 (PI: Ziurys), NSF MSIP Grant SVS-85009 / AST-1653228 (PI: Marrone), NSF CAREER Grant AST-1653228 (PI: Marrone), and a PIRE Grant DISE-1743747 (PI: Psaltis). I. Jim{\'e}nez-Serra and A. Meg{\'i}as acknowledge funding from the ERC grant OPENS (project number 101125858) funded by the European Union, and from the grant PID2022-136814NB-I00 funded by the Spanish Ministry of Science, Innovation and Universities / State Agency of Research, MCIU/AEI/10.13039/501100011033 and by "ERDF/EU". A. Taillard acknowledge funding from the European Research Council under the European Union's Horizon 2022 research and innovation program (grant agreement No. 101096293 SUL4LIFE).

\bibliographystyle{aa}
\bibliography{AA5860425}

\appendix

\onecolumn

\section{Observed Transitions}\label{obstran}

The observed c-C$_{3}$H$_{2}$, c-C$_{3}$HD and c-C$_{3}$D$_{2}$ transitions are presented in Tables \ref{tcc3h2}, \ref{tcc3hd} and \ref{tcc3d2}, respectively. The peak temperature ($T_{\rm MB}$), position ($V_{LSR}$), width (FWHM), rms and signal-to-noise ratio (SNR), calculated with the peak intensity, of the lines are given. The transitions that could not be fitted with pyspeckit are marked with an asterisk (*). For these transitions we provide the peak temperature, derived from the integrated area and an assumed FWHM value which is also indicated.\

\begin{longtable}{llccccc}
\caption{Observed c-C$_{3}$H$_{2}$ transitions.}\\
\hline\hline
Core & Transition & $T_{\rm MB}$ & $V_{LSR}$ & FWHM & rms & SNR \\
 & & (mK) & (km s$^{-1}$) & (km s$^{-1}$) & (mK) & \\
\hline\hline
264 & 3$_{21}$ - 3$_{12}$ & 45 & 7.996 (0.006) & 0.622 (0.015) & 6 & 7 \\
& 2$_{11}$ - 2$_{02}$ & 196 & 7.915 (0.001) & 0.508 (0.003) & 6 & 34 \\
& 2$_{20}$ - 1$_{11}$ & 152 & 7.916 (0.001) & 0.481 (0.003) & 19 & 8 \\
& 4$_{04}$ - 3$_{13}$ & 131 & 7.928 (0.001) & 0.460 (0.003) & 18 & 7 \\
& 3$_{12}$ - 2$_{21}$ & 338 & 7.937 (0.000) & 0.502 (0.001) & 39 & 9 \\
& 4$_{14}$ - 3$_{03}$ & 381 & 8.028 (0.000) & 0.488 (0.001) & 20 & 19 \\
\hline
317 & 3$_{21}$ - 3$_{12}$ & 23 & 8.036 (0.012) & 0.824 (0.029) & 5 & 4 \\
& 2$_{11}$ - 2$_{02}$ & 64 & 8.253 (0.004) & 0.619 (0.010) & 7 & 9 \\
& 2$_{20}$ - 1$_{11}$ & 59 & 8.226 (0.004) & 0.766 (0.009) & 12 & 5 \\
& 4$_{04}$ - 3$_{13}$ & 60 & 8.319 (0.003) & 0.671 (0.008) & 11 & 5 \\
& 3$_{12}$ - 2$_{21}$ & 150 & 8.262 (0.001) & 0.510 (0.003) & 23 & 7 \\
& 4$_{14}$ - 3$_{03}$ & 168 & 8.397 (0.001) & 0.637 (0.003) & 18 & 9 \\
\hline
321 & 3$_{21}$ - 3$_{12}$ & 38 & 8.430 (0.007) & 0.572 (0.016) & 4 & 9 \\
& 2$_{11}$ - 2$_{02}$ & 65 & 8.371 (0.005) & 0.781 (0.012) & 6 & 10 \\
& 2$_{20}$ - 1$_{11}$ & 71 & 8.542 (0.003) & 0.659 (0.006) & 15 & 5 \\
& 4$_{04}$ - 3$_{13}$ & 121 & 8.629 (0.002) & 0.664 (0.004) & 21 & 6 \\
& 3$_{12}$ - 2$_{21}$ & 260 & 8.560 (0.001) & 0.437 (0.001) & 33 & 8 \\
& 4$_{14}$ - 3$_{03}$ & 325 & 8.576 (0.001) & 0.591 (0.001) & 21 & 15 \\
\hline
326 & 3$_{21}$ - 3$_{12}$ & 64 & 7.958 (0.005) & 0.801 (0.011) & 3 & 20 \\
& 2$_{11}$ - 2$_{02}$ & 120 & 7.962 (0.003) & 0.935 (0.007) & 6 & 21 \\
& 2$_{20}$ - 1$_{11}$ & 160 & 8.188 (0.001) & 0.604 (0.003) & 18 & 9 \\
& 4$_{04}$ - 3$_{13}$ & 170 & 8.134 (0.001) & 0.854 (0.003) & 27 & 6 \\
& 3$_{12}$ - 2$_{21}$ & 364 & 8.151 (0.001) & 0.651 (0.001) & 37 & 10 \\
& 4$_{14}$ - 3$_{03}$ & 502 & 8.149 (0.000) & 0.846 (0.001) & 22 & 23 \\
\hline
413 & 3$_{21}$ - 3$_{12}$ & 94 & 7.773 (0.003) & 0.620 (0.007) & 7 & 14 \\
& 2$_{11}$ - 2$_{02}$ & 508 & 7.785 (0.000) & 0.534 (0.001) & 7 & 76 \\
& 2$_{20}$ - 1$_{11}$ & 353 & 7.786 (0.000) & 0.396 (0.001) & 27 & 13 \\
& 4$_{04}$ - 3$_{13}$ & 432 & 7.840 (0.000) & 0.346 (0.001) & 27 & 16 \\
& 3$_{12}$ - 2$_{21}$ & 806 & 7.797 (0.000) & 0.414 (0.000) & 49 & 17 \\
& 4$_{14}$ - 3$_{03}$ & 717 & 7.845 (0.000) & 0.437 (0.000) & 32 & 22 \\
\hline
504 & 3$_{21}$ - 3$_{12}$ & 42 & 6.775 (0.005) & 0.454 (0.013) & 2 & 18 \\
& 2$_{11}$ - 2$_{02}$ & 220 & 6.742 (0.001) & 0.476 (0.002) & 4 & 49 \\
& 2$_{20}$ - 1$_{11}$ & 154 & 6.749 (0.001) & 0.257 (0.001) & 34 & 5 \\
& 4$_{04}$ - 3$_{13}$ & 180 & 6.793 (0.001) & 0.242 (0.001) & 22 & 8 \\
& 3$_{12}$ - 2$_{21}$ & 398 & 6.791 (0.000) & 0.307 (0.001) & 41 & 10 \\
& 4$_{14}$ - 3$_{03}$ & 355 & 6.816 (0.000) & 0.279 (0.001) & 32 & 11 \\
\hline
615 & 3$_{21}$ - 3$_{12}$ & 20 & 7.956 (0.011) & 0.449 (0.025) & 6 & 3 \\
& 2$_{11}$ - 2$_{02}$ & 178 & 8.106 (0.001) & 0.503 (0.003) & 5 & 33 \\
& 2$_{20}$ - 1$_{11}$ & 87 & 8.062 (0.002) & 0.474 (0.004) & 13 & 7 \\
& 4$_{04}$ - 3$_{13}$ & 82 & 8.084 (0.002) & 0.436 (0.004) & 11 & 7 \\
& 3$_{12}$ - 2$_{21}$ & 249 & 8.063 (0.001) & 0.402 (0.001) & 39 & 6 \\
& 4$_{14}$ - 3$_{03}$ & 199 & 8.172 (0.001) & 0.383 (0.001) & 13 & 15 \\
\hline
627 & 3$_{21}$ - 3$_{12}$ & 48 & 8.380 (0.006) & 0.338 (0.009) & 6 & 8 \\
& 2$_{11}$ - 2$_{02}$ & 178 & 8.460 (0.001) & 0.375 (0.003) & 8 & 21 \\
& 2$_{20}$ - 1$_{11}$ & 139 & 8.394 (0.001) & 0.321 (0.002) & 21 & 7 \\
& 4$_{04}$ - 3$_{13}$ & 155 & 8.388 (0.001) & 0.332 (0.002) & 30 & 5 \\
& 3$_{12}$ - 2$_{21}$ & 285 & 8.411 (0.000) & 0.290 (0.001) & 48 & 6 \\
& 4$_{14}$ - 3$_{03}$ & 247 & 8.471 (0.000) & 0.365 (0.001) & 23 & 11 \\

\newpage
\caption[]{Observed c-C$_{3}$H$_{2}$ transitions (continued).} \\
\hline\hline
Core & Transition & $T_{\rm MB}$ & $V_{LSR}$ & FWHM & rms & SNR \\
 & & (mK) & (km s$^{-1}$) & (km s$^{-1}$) & (mK) & \\
\hline\hline
709 & 3$_{21}$ - 3$_{12}$ & 30 & 8.519 (0.010) & 0.797 (0.024) & 6 & 5 \\
& 2$_{11}$ - 2$_{02}$ & 81 & 8.589 (0.004) & 0.981 (0.009) & 6 & 13 \\
& 2$_{20}$ - 1$_{11}$ & 78 & 8.487 (0.003) & 0.760 (0.006) & 24 & 3 \\
& 4$_{04}$ - 3$_{13}$ & 85 & 8.547 (0.002) & 0.860 (0.006) & 16 & 5 \\
& 3$_{12}$ - 2$_{21}$ & 199 & 8.513 (0.001) & 0.842 (0.003) & 31 & 6 \\
& 4$_{14}$ - 3$_{03}$ & 244 & 8.538 (0.001) & 0.915 (0.003) & 14 & 18 \\
\hline
715 & 3$_{21}$ - 3$_{12}$ & 30 & 8.826 (0.011) & 0.782 (0.026) & 4 & 8 \\
& 2$_{11}$ - 2$_{02}$ & 119 & 8.923 (0.002) & 0.624 (0.006) & 5 & 22 \\
& 2$_{20}$ - 1$_{11}$ & 82 & 8.913 (0.003) & 0.632 (0.006) & 14 & 6 \\
& 4$_{04}$ - 3$_{13}$ & 83 & 9.039 (0.001) & 0.239 (0.003) & 13 & 6 \\
& 3$_{12}$ - 2$_{21}$ & 201 & 8.917 (0.001) & 0.621 (0.002) & 55 & 4 \\
& 4$_{14}$ - 3$_{03}$ & 262 & 9.085 (0.000) & 0.230 (0.001) & 12 & 21 \\
\hline
746 & 3$_{21}$ - 3$_{12}$ & 48 & 8.923 (0.006) & 0.362 (0.008) & 7 & 7 \\
& 2$_{11}$ - 2$_{02}$ & 190 & 8.892 (0.001) & 0.484 (0.003) & 8 & 24 \\
& 2$_{20}$ - 1$_{11}$ & 95 & 8.927 (0.001) & 0.389 (0.003) & 18 & 5 \\
& 4$_{04}$ - 3$_{13}$ & 111 & 9.028 (0.001) & 0.252 (0.002) & 18 & 6 \\
& 3$_{12}$ - 2$_{21}$ & 289 & 8.992 (0.000) & 0.370 (0.001) & 33 & 9 \\
& 4$_{14}$ - 3$_{03}$ & 260 & 9.039 (0.000) & 0.321 (0.001) & 28 & 9 \\
\hline
768 & 3$_{21}$ - 3$_{12}$ & 76 & 8.906 (0.003) & 0.557 (0.008) & 4 & 21 \\
& 2$_{11}$ - 2$_{02}$ & 265 & 8.931 (0.001) & 0.615 (0.002) & 3 & 86 \\
& 2$_{20}$ - 1$_{11}$ & 267 & 9.004 (0.000) & 0.344 (0.001) & 47 & 6 \\
& 4$_{04}$ - 3$_{13}$ & 321 & 9.138 (0.000) & 0.379 (0.001) & 44 & 7 \\
& 3$_{12}$ - 2$_{21}$ & 523 & 9.049 (0.000) & 0.420 (0.001) & 60 & 9 \\
& 4$_{14}$ - 3$_{03}$ & 610 & 9.105 (0.000) & 0.461 (0.001) & 47 & 13 \\
\hline
799 & $^{*}$3$_{21}$ - 3$_{12}$ & 41 &  & 0.025 & 5 & 8  \\
& 2$_{11}$ - 2$_{02}$ & 259 & 10.146 (0.001) & 0.386 (0.002) & 7 & 39 \\
& 2$_{20}$ - 1$_{11}$ & 170 & 10.213 (0.001) & 0.250 (0.001) & 16 & 10 \\
& 4$_{04}$ - 3$_{13}$ & 74 & 10.240 (0.001) & 0.246 (0.003) & 17 & 4 \\
& 3$_{12}$ - 2$_{21}$ & 367 & 10.219 (0.000) & 0.339 (0.001) & 54 & 7 \\
& 4$_{14}$ - 3$_{03}$ & 363 & 10.277 (0.000) & 0.247 (0.001) & 26 & 14 \\
\hline
800 & 3$_{21}$ - 3$_{12}$ & 89 & 10.310 (0.003) & 0.564 (0.007) & 6 & 16 \\
& 2$_{11}$ - 2$_{02}$ & 310 & 10.316 (0.001) & 0.572 (0.002) & 5 & 60 \\
& 2$_{20}$ - 1$_{11}$ & 185 & 10.421 (0.001) & 0.735 (0.002) & 29 & 6 \\
& 4$_{04}$ - 3$_{13}$ & 234 & 10.528 (0.001) & 0.470 (0.001) & 28 & 8 \\
& 3$_{12}$ - 2$_{21}$ & 572 & 10.463 (0.000) & 0.429 (0.001) & 47 & 12 \\
& 4$_{14}$ - 3$_{03}$ & 503 & 10.516 (0.000) & 0.494 (0.001) & 32 & 16 \\
\hline
\label{tcc3h2}
\end{longtable}

\newpage

\begin{longtable}{llccccc}
\caption{Observed c-C$_{3}$HD transitions.}\\
\hline\hline
Core & Transition & $T_{\rm MB}$ & $V_{LSR}$ & FWHM & rms & SNR \\
 & & (mK) & (km s$^{-1}$) & (km s$^{-1}$) & (mK) & \\
\hline\hline
264 & 2$_{11}$ - 2$_{02}$ & 21 & 7.935 (0.013) & 0.590 (0.030) & 2 & 11 \\
& 1$_{11}$ - 0$_{00}$ & 44 & 8.011 (0.006) & 0.677 (0.014) & 12 & 4 \\
& 4$_{14}$ - 3$_{03}$ & 43 & 7.985 (0.004) & 0.514 (0.009) & 6 & 7 \\
& 2$_{20}$ - 1$_{11}$ & 34 & 8.035 (0.005) & 0.511 (0.011) & 6 & 6 \\
\hline
317 & 2$_{11}$ - 2$_{02}$ & 14 & 8.280 (0.020) & 0.605 (0.048) & 3 & 5 \\
& 1$_{11}$ - 0$_{00}$ & 38 & 8.210 (0.006) & 0.465 (0.013) & 9 & 4 \\
& 4$_{14}$ - 3$_{03}$ & 33 & 8.315 (0.003) & 0.273 (0.008) & 9 & 4 \\
& 2$_{20}$ - 1$_{11}$ & 28 & 8.459 (0.005) & 0.430 (0.012) & 8 & 3 \\
\hline
321 & 2$_{11}$ - 2$_{02}$ & 15 & 8.539 (0.019) & 0.562 (0.043) & 3 & 6 \\
& 1$_{11}$ - 0$_{00}$ & 23 & 8.366 (0.009) & 0.540 (0.022) & 7 & 4 \\
& 4$_{14}$ - 3$_{03}$ & 57 & 8.523 (0.003) & 0.489 (0.007) & 11 & 5 \\
& 2$_{20}$ - 1$_{11}$ & 162 & 8.508 (0.003) & 0.439 (0.008) & 11 & 15 \\
\hline
326 & 2$_{11}$ - 2$_{02}$ & 24 & 7.998 (0.015) & 0.979 (0.036) & 2 & 12 \\
& 1$_{01}$ - 0$_{00}$ & 10 & 7.850 (0.030) & 0.807 (0.071) & 2 & 4 \\
& 1$_{11}$ - 0$_{00}$ & 74 & 8.061 (0.004) & 0.652 (0.009) & 6 & 12 \\
& 4$_{14}$ - 3$_{03}$ & 71 & 8.087 (0.003) & 0.936 (0.007) & 13 & 5 \\
& 2$_{20}$ - 1$_{11}$ & 173 & 8.157 (0.004) & 0.969 (0.009) & 13 & 13 \\
\hline
413 & 2$_{11}$ - 2$_{02}$ & 77 & 7.816 (0.004) & 0.635 (0.009) & 4 & 22 \\
& 1$_{01}$ - 0$_{00}$ & 27 & 7.896 (0.013) & 0.855 (0.030) & 4 & 7 \\
& 1$_{11}$ - 0$_{00}$ & 240 & 7.874 (0.001) & 0.627 (0.003) & 8 & 31 \\
& 4$_{14}$ - 3$_{03}$ & 141 & 7.779 (0.001) & 0.285 (0.002) & 13 & 11 \\
& 2$_{20}$ - 1$_{11}$ & 146 & 7.823 (0.001) & 0.342 (0.002) & 12 & 13 \\
\hline
504 & 2$_{11}$ - 2$_{02}$ & 46 & 6.783 (0.006) & 0.629 (0.015) & 2 & 19 \\
& 1$_{01}$ - 0$_{00}$ & 21 & 6.718 (0.013) & 0.701 (0.032) & 3 & 6 \\
& 1$_{11}$ - 0$_{00}$ & 150 & 6.788 (0.002) & 0.573 (0.004) & 8 & 20 \\
& 4$_{14}$ - 3$_{03}$ & 90 & 6.703 (0.001) & 0.249 (0.002) & 13 & 7 \\
& 2$_{20}$ - 1$_{11}$ & 91 & 6.755 (0.001) & 0.319 (0.003) & 11 & 9 \\
\hline
615 & 2$_{11}$ - 2$_{02}$ & 23 & 8.080 (0.011) & 0.584 (0.027) & 2 & 12 \\
& 1$_{11}$ - 0$_{00}$ & 82 & 8.175 (0.003) & 0.504 (0.006) & 11 & 8 \\
& 4$_{14}$ - 3$_{03}$ & 17 & 8.051 (0.012) & 0.559 (0.028) & 6 & 3 \\
& 2$_{20}$ - 1$_{11}$ & 28 & 8.158 (0.004) & 0.348 (0.010) & 7 & 4 \\
\hline
627 & 2$_{11}$ - 2$_{02}$ & 26 & 8.348 (0.011) & 0.535 (0.023) & 3 & 8 \\
& 1$_{11}$ - 0$_{00}$ & 118 & 8.507 (0.002) & 0.525 (0.005) & 10 & 12 \\
& 4$_{14}$ - 3$_{03}$ & 39 & 8.331 (0.003) & 0.259 (0.006) & 6 & 7 \\
& 2$_{20}$ - 1$_{11}$ & 55 & 8.400 (0.002) & 0.308 (0.005) & 6 & 9 \\
\hline
709 & 2$_{11}$ - 2$_{02}$ & 18 & 8.521 (0.017) & 0.747 (0.040) & 3 & 5 \\
& 1$_{11}$ - 0$_{00}$ & 46 & 8.390 (0.006) & 0.680 (0.014) & 8 & 6 \\
& 4$_{14}$ - 3$_{03}$ & 26 & 8.445 (0.007) & 0.631 (0.016) & 5 & 6 \\
& 2$_{20}$ - 1$_{11}$ & 25 & 8.441 (0.008) & 0.670 (0.018) & 5 & 5 \\
\hline
715 & 2$_{11}$ - 2$_{02}$ & 15 & 8.774 (0.026) & 0.826 (0.060) & 2 & 7 \\
& 1$_{11}$ - 0$_{00}$ & 38 & 8.951 (0.008) & 0.698 (0.018) & 10 & 4 \\
& 4$_{14}$ - 3$_{03}$ & 24 & 9.008 (0.004) & 0.212 (0.009) & 4 & 5 \\
& 2$_{20}$ - 1$_{11}$ & 27 & 9.055 (0.004) & 0.263 (0.010) & 5 & 5 \\
\hline
746 & 2$_{11}$ - 2$_{02}$ & 34 & 8.952 (0.008) & 0.514 (0.018) & 3 & 12 \\
& 1$_{11}$ - 0$_{00}$ & 122 & 8.966 (0.002) & 0.520 (0.005) & 8 & 15 \\
& 4$_{14}$ - 3$_{03}$ & 43 & 9.028 (0.006) & 0.748 (0.014) & 7 & 6 \\
& 2$_{20}$ - 1$_{11}$ & 35 & 8.995 (0.004) & 0.407 (0.010) & 8 & 4 \\
\hline
768 & 2$_{11}$ - 2$_{02}$ & 19 & 8.914 (0.014) & 0.495 (0.028) & 2 & 9 \\
& 1$_{11}$ - 0$_{00}$ & 62 & 8.978 (0.004) & 0.561 (0.009) & 4 & 14 \\
& 4$_{14}$ - 3$_{03}$ & 37 & 9.047 (0.003) & 0.336 (0.007) & 7 & 6 \\
& 2$_{20}$ - 1$_{11}$ & 45 & 9.108 (0.003) & 0.426 (0.007) & 5 & 8 \\
\hline

\newpage

\caption[]{Observed c-C$_{3}$HD transitions (continued).}\\
\hline\hline
Core & Transition & $T_{\rm MB}$ & $V_{LSR}$ & FWHM & rms & SNR \\
 & & (mK) & (km s$^{-1}$) & (km s$^{-1}$) & (mK) & \\
\hline\hline
799 & 2$_{11}$ - 2$_{02}$ & 39 & 10.135 (0.008) & 0.385 (0.019) & 3 & 12 \\
& 1$_{11}$ - 0$_{00}$ & 138 & 10.192 (0.002) & 0.511 (0.004) & 12 & 11 \\
& 4$_{14}$ - 3$_{03}$ & 46 & 10.195 (0.002) & 0.217 (0.005) & 5 & 8 \\
& 2$_{20}$ - 1$_{11}$ & 58 & 10.274 (0.002) & 0.263 (0.004) & 4 & 15 \\
\hline
800 & 2$_{11}$ - 2$_{02}$ & 68 & 10.336 (0.005) & 0.716 (0.011) & 3 & 22 \\
& 1$_{01}$ - 0$_{00}$ & 27 & 10.322 (0.012) & 0.821 (0.028) & 4 & 6 \\
& 1$_{11}$ - 0$_{00}$ & 233 & 10.382 (0.001) & 0.633 (0.003) & 10 & 24 \\
& 4$_{14}$ - 3$_{03}$ & 93 & 10.418 (0.001) & 0.491 (0.003) & 9 & 10 \\
& 2$_{20}$ - 1$_{11}$ & 106 & 10.476 (0.001) & 0.522 (0.004) & 11 & 10 \\
\hline
L1448 & 3$_{03}$ - 2$_{02}$ & 33 & 4.676 (0.005) & 0.353 (0.011) & 4 & 8 \\
\hline
\label{tcc3hd}
\end{longtable}

\begin{longtable}{llccccc}
\caption{Observed c-C$_{3}$D$_{2}$ transitions.}\\
\hline\hline
Core & Transition & $T_{\rm MB}$ & $V_{LSR}$ & FWHM & rms & SNR \\
 & & (mK) & (km s$^{-1}$) & (km s$^{-1}$) & (mK) & \\
\hline\hline
321 & 1$_{11}$ - 0$_{00}$ & 13 & 8.537 (0.019) & 0.586 (0.045) & 4 & 3 \\
\hline
413 & 1$_{11}$ - 0$_{00}$ & 23 & 7.904 (0.009) & 0.504 (0.024) & 4 & 6 \\
\hline
504 & 2$_{11}$ - 2$_{02}$ & 7 & 6.690 (0.051) & 0.836 (0.119) & 2 & 4 \\
& 1$_{11}$ - 0$_{00}$ & 34 & 6.867 (0.007) & 0.548 (0.018) & 4 & 9 \\
\hline
627 & $^{*}$ 1$_{11}$ - 0$_{00}$ & 35  &  & 0.25 & 5 &  7\\
\hline
715 & 1$_{11}$ - 0$_{00}$ & 13 & 9.082 (0.020) & 0.552 (0.046) & 4 & 4 \\
\hline
746 & 1$_{11}$ - 0$_{00}$ & 26 & 8.939 (0.008) & 0.475 (0.022) & 6 & 4 \\
\hline
799 & $^{*}$ 1$_{11}$ - 0$_{00}$ & 29  &  & 0.25 & 5 &  6\\
\hline
800 & 2$_{11}$ - 2$_{02}$ & 9 & 10.325 (0.035) & 0.876 (0.084) & 2 & 4 \\
& 1$_{11}$ - 0$_{00}$ & 42 & 10 (0.005) & 0.479 (0.013) & 6 & 7 \\
\hline
L1448 & 3$_{03}$ - 2$_{12}$ & 29 & 4.965 (0.005) & 0.300 (0.011) & 4 & 7 \\
\hline
\label{tcc3d2}
\end{longtable}

\twocolumn

\section{Column Density calculation}\label{columndensity}

\begin{figure}[H]
\centering
\includegraphics[width=7cm]{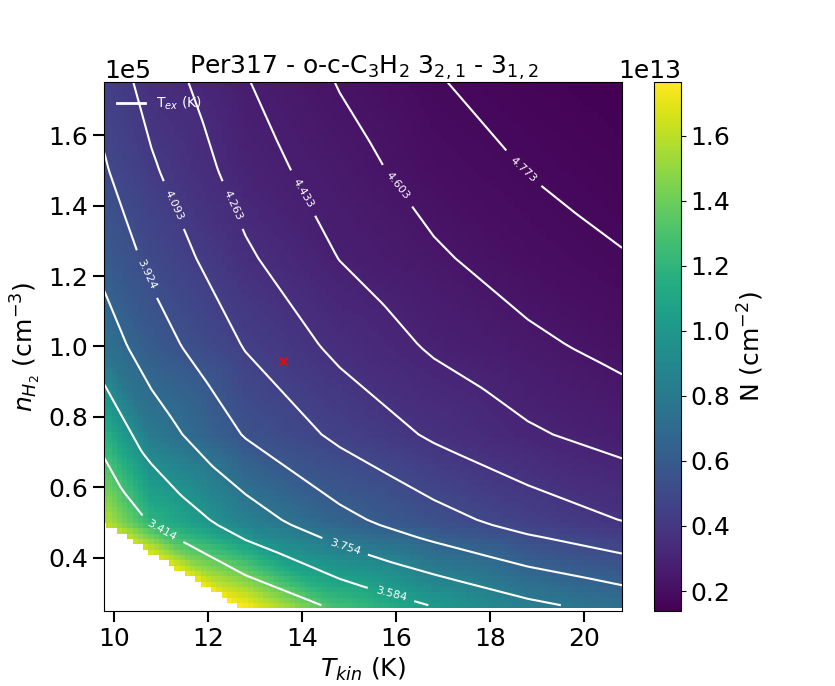}
\caption{Column density of the ortho-c-C$_{3}$H$_{2}$ 3$_{2,1}$ - 3$_{1,2}$ transition of Per317 plotted with a colour gradient for a range of $n_{\rm H_2}$ and $T_{\rm kin}$. The contour white lines indicate the excitation temperature of the transition, $T_{\rm ex}$, and the red cross marks the $n_{\rm H_2}$ and $T_{\rm kin}$ values derived in \cite{pezzuto:21}. }
\label{grido}
\end{figure}

\begin{figure}[H]
\centering
\includegraphics[width=7cm]{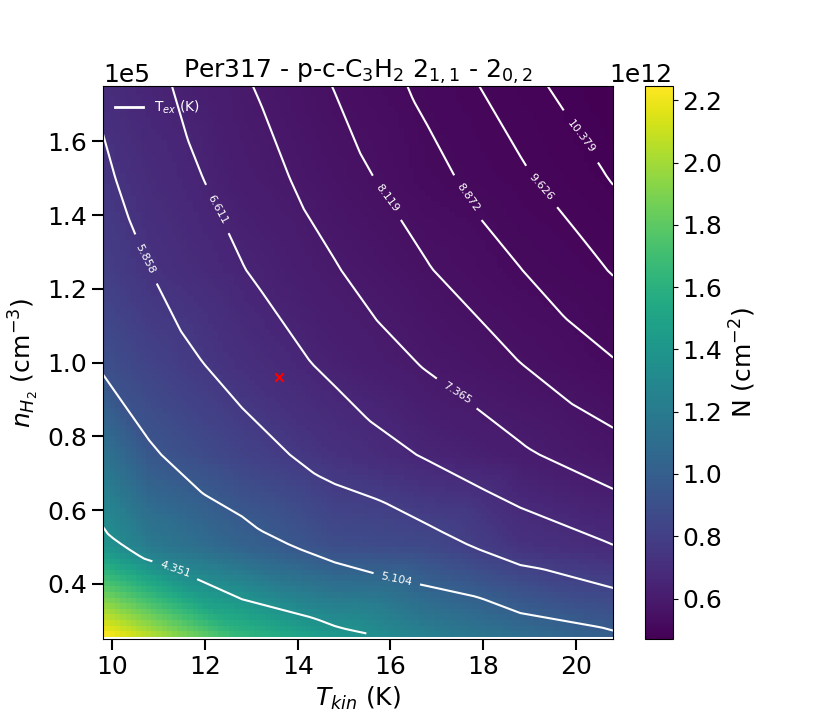}
\caption{Column density of the para-c-C$_{3}$H$_{2}$ 2$_{1,1}$ - 2$_{0,2}$ transition of Per317 plotted with a colour gradient for a range of $n_{\rm H_2}$ and $T_{\rm kin}$. The contour white lines indicate the excitation temperature of the transition, $T_{\rm ex}$, and the red cross marks the $n_{\rm H_2}$ and $T_{\rm kin}$ values derived in \cite{pezzuto:21}. }
\label{grid}
\end{figure}

With the aim of testing the viability of using fixed values for $n_{\rm H_2}$ and $T_{\rm kin}$ to compute the column density of c-C$_{3}$H$_{2}$ and its isotopologues with RADEX, we computed the column density for two transitions for a grid of $n_{\rm H_2}$ and $T_{\rm kin}$ values. An example core was selected, Per317, and the column density for the ortho-c-C$_{3}$H$_{2}$ 3$_{2,1}$ - 3$_{1,2}$ and para-c-C$_{3}$H$_{2}$ 2$_{1,1}$ - 2$_{0,2}$ transitions were calculated for a $n_{\rm H_2}$ range of 0.25 - 1.75$\times$10$^{5}$ cm$^{-3}$ and for a $T_{\rm kin}$ range of 10 - 21 K (Figures \ref{grido} and \ref{grid}, respectively). The obtained column density values within the 10\% observational uncertainty of the constrained $n_{\rm H_2}$ and $T_{\rm kin}$ values taken from \cite{scibelli:24}, varies by less than a factor of 2. Thus, the approach of fixing the $n_{\rm H_2}$ and $T_{\rm kin}$ values while taking into account a 10\% observational uncertainty, is a reasonable assumption to derive the column densities of c-C$_{3}$H$_{2}$ and its isotopologues.\

\subsection{Output $n_{\rm H_2}$ and $T_{\rm kin}$}\label{constrained}

We report the constrained $n_{\rm H_2}$ and $T_{\rm kin}$ by RADEX for each of the molecules: orhto-c-C$_{3}$H$_{2}$, para-c-C$_{3}$H$_{2}$, c-C$_{3}$HD and ortho-c-C$_{3}$D$_{2}$ (Tables \ref{occ3h2phys}, \ref{pcc3h2phys}, \ref{cc3hdphys} and  \ref{occ3d2phys}, respectively). As seen above, the column density does not change significantly within the 10\% error assumed for $n_{\rm H_2}$ and $T_{\rm kin}$. That is why the standard deviation of the volume density and kinetic temperature range within this 10\% uncertainty, and thus why the values are not tightly constrained.\

\begin{table}[H]
\begin{center}
\caption{Constrained $n_{\rm H_2}$ and $T_{\rm kin}$ from the RADEX fitting of ortho-c-C$_{3}$H$_{2}$.}
\begin{tabular}{c c c}

\hline\hline
Core & $n_{\rm H_2}$ & $T_{\rm kin}$ \\
 &  (cm$^{-3}$) & (K)  \\
\hline\hline
264 & $7.5 \pm 0.7\times10^{4}$ & $11 \pm 1$ \\
317 & $9.4 \pm 0.9\times10^{4}$ & $14 \pm 1$ \\
321 & $1.3 \pm 0.1\times10^{5}$ & $12 \pm 1$ \\
326 & $2.9 \pm 0.3\times10^{5}$ & $12 \pm 1$ \\
413 & $4.2 \pm 0.4\times10^{4}$ & $11 \pm 1$ \\
504 & $6.5 \pm 0.7\times10^{4}$ & $10 \pm 1$ \\
615 & $3.5 \pm 0.3\times10^{4}$ & $10 \pm 1$ \\
627 & $3.2 \pm 0.3\times10^{4}$ & $12 \pm 1$ \\
709 & $5.1 \pm 0.5\times10^{4}$ & $14 \pm 1$ \\
715 & $4.9 \pm 0.5\times10^{4}$ & $15 \pm 2$ \\
746 & $4.9 \pm 0.5\times10^{4}$ & $11 \pm 1$ \\
768 & $4.9 \pm 0.5\times10^{4}$ & $11 \pm 1$ \\
799 & $3.3 \pm 0.3\times10^{4}$ & $10 \pm 1$ \\
800 & $6.4 \pm 0.7\times10^{4}$ & $12 \pm 1$ \\
\hline
\label{occ3h2phys}
\end{tabular}
\end{center}
\end{table}

\begin{table}[H]
\begin{center}
\caption{Constrained $n_{\rm H_2}$ and $T_{\rm kin}$ from the RADEX fitting of para-c-C$_{3}$H$_{2}$.}
\begin{tabular}{c c c}

\hline\hline
Core & $n_{\rm H_2}$ & $T_{\rm kin}$ \\
 &  (cm$^{-3}$) & (K)  \\
\hline\hline
264 & $7.5 \pm 0.7\times10^{4}$ & $11 \pm 1$ \\
317 & $9.7 \pm 0.9\times10^{4}$ & $13 \pm 1$ \\
321 & $1.3 \pm 0.1\times10^{5}$ & $13 \pm 1$ \\
326 & $2.6 \pm 0.2\times10^{5}$ & $12 \pm 1$ \\
413 & $4.2 \pm 0.4\times10^{4}$ & $10 \pm 1$ \\
504 & $6.5 \pm 0.6\times10^{4}$ & $10 \pm 1$ \\
615 & $3.4 \pm 0.3\times10^{4}$ & $10 \pm 1$ \\
627 & $3.2 \pm 0.3\times10^{4}$ & $12 \pm 1$ \\
709 & $5.1 \pm 0.6\times10^{4}$ & $13 \pm 1$ \\
715 & $4.9 \pm 0.5\times10^{4}$ & $15 \pm 1$ \\
746 & $4.9 \pm 0.5\times10^{4}$ & $11 \pm 1$ \\
768 & $4.9 \pm 0.5\times10^{4}$ & $11 \pm 1$ \\
799 & $3.3 \pm 0.3\times10^{4}$ & $10 \pm 1$ \\
800 & $6.4 \pm 0.6\times10^{4}$ & $11 \pm 1$ \\
\hline
\label{pcc3h2phys}
\end{tabular}
\end{center}
\end{table}

\begin{table}[H]
\begin{center}
\caption{Constrained $n_{\rm H_2}$ and $T_{\rm kin}$ from the RADEX fitting of c-C$_{3}$HD.}
\begin{tabular}{c c c}

\hline\hline
Core & $n_{\rm H_2}$ & $T_{\rm kin}$ \\
 &  (cm$^{-3}$) & (K)  \\
\hline\hline
264 & $7.5 \pm 0.8\times10^{4}$ & $11 \pm 1$ \\
317 & $9.6 \pm 1.0\times10^{4}$ & $14 \pm 1$ \\
321 & $1.3 \pm 0.1\times10^{5}$ & $13 \pm 1$ \\
326 & $2.8 \pm 0.3\times10^{5}$ & $12 \pm 1$ \\
413 & $4.2 \pm 0.4\times10^{4}$ & $10 \pm 1$ \\
504 & $6.5 \pm 0.7\times10^{4}$ & $10 \pm 1$ \\
615 & $3.5 \pm 0.3\times10^{4}$ & $10 \pm 1$ \\
627 & $3.2 \pm 0.3\times10^{4}$ & $12 \pm 1$ \\
709 & $5.1 \pm 0.5\times10^{4}$ & $14 \pm 1$ \\
715 & $4.9 \pm 0.4\times10^{4}$ & $15 \pm 1$ \\
746 & $4.8 \pm 0.5\times10^{4}$ & $11 \pm 1$ \\
768 & $4.8 \pm 0.5\times10^{4}$ & $11 \pm 1$ \\
799 & $3.3 \pm 0.4\times10^{4}$ & $10 \pm 1$ \\
800 & $6.3 \pm 0.6\times10^{4}$ & $12 \pm 1$ \\
L448 & $5.7 \pm 0.6\times10^{4}$ & $15 \pm 1$ \\
\hline
\label{cc3hdphys}
\end{tabular}
\end{center}
\end{table}

\begin{table}[H]
\begin{center}
\caption{Constrained $n_{\rm H_2}$ and $T_{\rm kin}$ from the RADEX fitting of ortho-c-C$_{3}$D$_{2}$.}
\begin{tabular}{c c c}

\hline\hline
Core & $n_{\rm H_2}$ & $T_{\rm kin}$ \\
 &  (cm$^{-3}$) & (K)  \\
\hline\hline
321 & $(1.3 \pm 0.1)\times10^{5}$ & $13 \pm 1$ \\
413 & $(4.2 \pm 0.4)\times10^{4}$ & $11 \pm 1$ \\
504 & $(6.5 \pm 0.7)\times10^{4}$ & $10 \pm 1$ \\
627 & $(3.2 \pm 0.3)\times10^{4}$ & $12 \pm 1$ \\
715 & $(4.9 \pm 0.5)\times10^{4}$ & $15 \pm 2$ \\
746 & $(4.9 \pm 0.5)\times10^{4}$ & $11 \pm 1$ \\
799 & $(3.3 \pm 0.3)\times10^{4}$ & $10 \pm 1$ \\
800 & $(6.4 \pm 0.7)\times10^{4}$ & $12 \pm 1$ \\

\hline
\label{occ3d2phys}
\end{tabular}
\end{center}
\end{table}

\newpage

\section{Fractional Abundance}\label{abundance}

\begin{figure*}[hb]
\centering
\includegraphics[width=\textwidth]{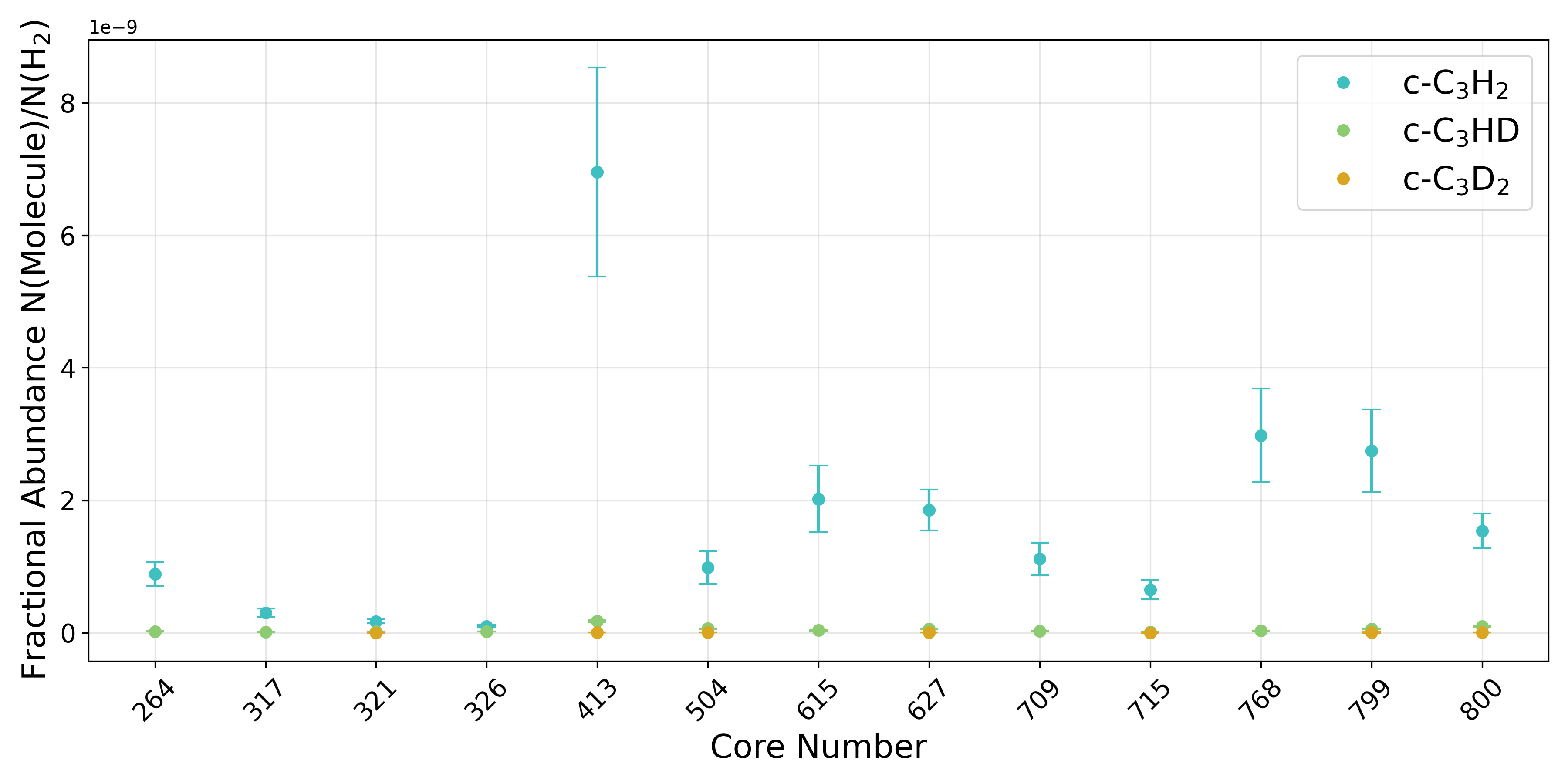}
\caption{Abundance of c-C$_{3}$H$_{2}$, c-C$_{3}$HD and c-C$_{3}$D$_{2}$ with respect to molecular hydrogen in blue, green and orange, respectively.}
\label{abun}
\end{figure*}

The fractional abundance of c-C$_{3}$H$_{2}$, c-C$_{3}$HD and c-C$_{3}$D$_{2}$ with respect to molecular hydrogen are plotted for the studied starless and prestellar cores in the Perseus Molecular Cloud (Figure \ref{abun}).\

\end{document}